\documentclass[10pt,a4]{article}
\usepackage{threeparttable}
\usepackage{amsmath,amssymb,amsthm}
\usepackage{color}
\usepackage{collcell}
\usepackage[english]{babel}
\usepackage{graphicx}
\usepackage{natbib}
\usepackage{subfig}
\usepackage{threeparttable}
\usepackage{multirow}

\newcommand{\R}{\mathbb R}
\newcommand{\s}{\mathcal S}
\newcommand{\de}{\mathrm d}

\newcommand\norm[1]{\left\lVert#1\right\rVert}
\newcommand{\lo}{\mathrm{low}}
\newcommand{\up}{\mathrm{upp}}
\usepackage{mathtools}

\usepackage{geometry}
\newtheorem{def.}{Definition}[section]
\newtheorem{examp.}{Example}[section]

\RequirePackage[colorlinks,citecolor=blue,urlcolor=blue]{hyperref}

\begin{document}

\title{
 Testing the first-order separability hypothesis for spatio-temporal point patterns
}
\date{}

\author{}
\maketitle
  
\begin{center}
  {\small\bf Mohammad Ghorbani }\\\thanks{Department of Mathematics and Mathematical Statistics, Ume\r{a}  University, Sweden\\E-mail: 
    mohammad.ghorbani@umu.se} \\

  {\small\bf Nafiseh Vafaei}\\
 \thanks{Department of Computer and Statistics Sciences, Faculty of Sciences, Mohaghegh Ardabili University, Ardabil, Iran\\E-mail: Nafiseh.Vafaei@uma.ac.ir}\\  
   {\small\bf Ji\v{r}\'{i} Dvo\v{r}\'{a}k}\\
 \thanks{Department of Probability and Mathematical Statistics, Faculty of Mathematics and Physics, Charles University, Prague, Czech Republic
 \\E-mail: dvorak@karlin.mff.cuni.cz}\\ 

  {\small\bf Mari Myllym\"{a}ki}\\
 \thanks{Natural Resources Institute Finland (Luke),  Helsinki, Finland.\\E-mail: mari.myllymaki@luke.fi}\\ 

\end{center}
\begin{abstract}
First-order separability of a spatio-temporal point process plays a fundamental role in the analysis of spatio-temporal point pattern data. While it is often a convenient assumption that simplifies the analysis greatly, existing non-separable structures should be accounted for in the model construction.
We propose three different tests to investigate this hypothesis as a step of preliminary data analysis. 
The first two tests are exact or asymptotically exact for Poisson processes.
The first test based on permutations and global envelopes allows us to detect at which spatial and temporal locations or lags the data deviate from the null hypothesis.
The second test is a simple and computationally cheap $\chi^2$-test.
The third test is based on statistical reconstruction method and can be generally applied for non-Poisson processes.
The performance of the first two tests is studied in a simulation study for Poisson and non-Poisson models. The third test is applied to the real data of the UK 2001 epidemic foot and mouth  disease.

\end{abstract}
\section{Introduction}
Spatio-temporal point process (STPP) models are increasingly used for modeling natural phenomena like disease incidences, sightings or births of a species, occurrences of fires, earthquakes, tsunamis, or volcanic eruptions \citep{schoenberg:brillinger:guttorp:02,diggle:13}.
In practical applications, modeling the joint distribution
of spatial locations and occurrence times of a spatio-temporal point pattern
are challenging. Therefore, to make some simplification, it is often assumed that the point process has a separable spatio-temporal intensity function.
In general, the strongest form of separability may be defined by the requirement that the distribution of a STPP is equal to the product of the distributions of the marginal processes in space and time. This form of separability is equivalent to the 
independence of the spatial and temporal components of the point process, and under this separability, the
spatial and temporal components can be modeled completely separately \citep{Benes:etal:10}. 
\citet[page 220]{diggle:13}  uses the term {\em no spatio-temporal interaction} for a point process with independent spatial and temporal components. Weaker notions of separability are characterized by the product form of moment measures.

In order to discover what type of models would be appropriate for the data and also to help in interpreting summary characteristics in preliminary data analysis, an important part 
for STPPs is to consider different separability hypotheses. Depending on whether the pattern
is considered to be a realisation of a stationary process or inhomogeneous, the first steps in the analysis of the  data and the interest to separability hypotheses typically differ.

If an STPP is considered homogeneous, 
the interest typically lies in the interaction between the points and the first step is the test of complete spatio-temporal randomness \citep{diggle:13, illian:penttinen:stoyan:stoyan:08}. If the presence of interaction is confirmed,
one of the crucial steps is to test the second-order separability, i.e.\ separability of the space-time pair correlation function \citep{Gabriel:Diggle:09, moeller:ghorbani:12,ghorbani:13,Moeller:etal:19}. If no evidence against the separability is found, the interactions in space and time can be inspected separately of each other. On the other hand, if the pattern is deduced to origin from a  non-separable process, then interactions must be considered in space-time. 

For an inhomogeneous pattern, the focus is typically first on its intensity function and the first-order separability, i.e.\ the separability of the space-time intensity function.
First-order separability is often regarded as a convenient working assumption \citep{Gabriel:Diggle:09, moeller:ghorbani:12, Moeller:etal:19}, because under the separability the inference about the quite complicated spatio-temporal model can typically be based on the properties of the lower dimensional spatial and temporal marginal processes that are easier to handle. 
However, obviously the separability does not hold always, and a test of separability can help to suggest suitable explanatory variables for the variations in the intensity. Namely, under separability, spatial and temporal covariates may be used, whereas otherwise, one should look for spatio-temporal covariates.
If an inhomogeneous pattern additionally exhibits clustering or regularity, it may be of interest to test for the separability of the second-order property once a good model for the intensity exists. Such a Monte Carlo test can be performed if the first-order separability holds \citep{diggle:13}.

The first- and second-order separability are not fully explored in the previous literature. So far, in the context of spatio-temporal marked point processes, for testing the hypothesis of the  separability of the marks and the spatial-temporal process, some test statistics based on the conditional intensity function have been proposed by \cite{Schoenberg:2004} and \cite{Diaz-Avalos:etal:13}. Recently, \cite{Isabel:2018} provided a test statistic based on the relative risk function and using regression methods. All these test statistics  were applied to wildfire data.
As another test statistic for the hypothesis of the first-order separability, \cite{Gonzalez:2019} used the ratio of the integrated intensity function at two disjoint spatial regions  
which should not vary by time 
and employed it in the analysis of tornado occurrences in the USA.
However, none of these test statistic allows one to discover reasons of non-separability, i.e., where and when the non-separability occurs.

The aim of this paper is to propose tests for the null hypothesis of spatio-temporal separability of intensity function.
We propose three different tests to investigate this hypothesis, namely, a permutation based test, a test based on stochastic reconstruction, and  the $\chi^2$-test. For the first two tests, our test statistics are simply based on the nonparametric estimates of the non-separable and separable  intensity functions.
As usual in spatial and spatio-temporal statistics, the distribution functions of the proposed test statistics are unknown and Monte Carlo methods are used to compare the observed value of the test statistic with the values obtained from the simulated samples under the null hypothesis. In general, in the analysis of spatio-temporal point patterns, the test statistics are often summary functions, as here as well, and pointwise envelopes have been used for model checking \citep{Gabriel:Diggle:09,moeller:ghorbani:12,moeller:ghorbani:13}. However, while the functions are inspected on an interval or a region, the pointwise envelopes control the type of error only locally  \citep[see e.g.][]{loosmore:ford:2006,Baddeley:etal:16, mari:etal:13}.
We utilize instead the global envelope tests \citep{mari:etal:13, MyllymakiMrkvicka2019} that control the global type I error and show how they can be useful also in testing the separability hypothesis by providing good power and graphical interpretation of the test results. 
In producing Monte Carlo samples under the first-order separability hypothesis, a simple permutation strategy, namely permutation of times of the point pattern, works well for Poisson processes. However, it breaks down the second-order structures of non-Poisson processes. For non-Poisson cases, we suggest instead a test based on the stochastic reconstruction procedure \citep{TscheschelStoyan2006,WiegandEtAl2013,KonasovaDvorak}, which can be used to produce Monte Carlo replications with the same interaction structure as the observed data and the same intensity functions of the spatial and temporal component process $X_{{\mathrm{space}}}$ and $X_{{\mathrm{time}}}$, respectively. 
The third test is a computationally cheap $\chi^2$-test, which is based on the cell counts. The permutation based test is exact and the $\chi^2$-test is asymptotically exact for Poisson processes. The random reconstruction based test is instead appropriate for non-Poisson processes, provided that the reconstruction procedure is performed carefully.

The rest of this paper is organized as follows. Section~\ref{sec:background} provides some background materials about STPPs including their first-order characteristics that will be used in the subsequent sections. 
The null hypothesis of first-order separability is defined in Section~\ref{sec:first-order}.
The permutation based test is introduced in Section \ref{sec:permutation-test}, including test statistics, permutation strategy and description on the use of the statistics calculated for the data and simulations in the global envelope and deviation tests.
Sections~\ref{sec:chisquare} and \ref{sec:SR} respectively express the $\chi^2$-test and stochastic reconstruction method for testing the first-order separability. 
Section~\ref{sec:performance} is devoted to two simulation studies, one for inhomogeneous Poisson processes in Section~\ref{sec:simexample} and another for log-Gaussian Cox processes in Section \ref{sec:LGCP}. These studies compare the performance of the proposed tests under various alternative hypotheses and also explain the graphical interpretation of the global envelope tests. 
We applied the tests to the UK 2001 foot and mouth disease data in Cumbria \citep{keelingetal:01, Diggle:06,diggle:13,ghorbani:13,moeller:ghorbani:12} in Section~\ref{sec:fmd}. The paper ends with a short discussion. Some visualisation tools including interactive plots, suitable for informal assessment of the first-order separability hypothesis, are presented at the accompanying website: \url{http://msekce.karlin.mff.cuni.cz/~dvorak/software/STseparability.html}.

\section{Assumptions and background}\label{sec:background}
A 
STPP 
$X$ with no overlapping points is a random countable subset  of a space $\s\subseteq\mathbb{R}^2\times \mathbb{R}$, where each point $(u, t) \in X$  indicates the location and time of occurrence of an event of interest.  In practice, $X$ is observed within a spatio-temporal window $W\times T$, where
$W\subset\mathbb R^2$ is a bounded region of
area $|W|>0$, $T\subset\mathbb R$ is a bounded time interval
 of length $|T|>0$, and $X\cap (W\times T)=\{(u_i,t_i),\
 i=1\ldots n\}$ is the observed data. 
We assume that $X$ has intensity function $\rho$.

Assuming that $X$ has an intensity
 function $\rho(\cdot)$,
the spatial component process $X_{{\mathrm{space}}}$ consisting of
the locations with times in $T$ and the temporal
component process $X_{{\mathrm{time}}}$ consisting of the times with locations in $W$, i.e.\  $X_{{\mathrm{space}}}=\{u:\,(u,t)\in X,\, t\in T\}$ and  $X_{{\mathrm{time}}}=\{t:\, (u,t)\in X,\, u\in W\}$,
are then well-defined point processes on $\mathbb R^2$ and $\mathbb R$, respectively, with well-defined intensity  
functions.
Following \cite{moeller:ghorbani:12}, we use the indices {\em space} and {\em time} for the respective functional summaries of the spatial and temporal components. The intensity functions of these components are given by 
\begin{align}
\label{eq:et:compnts}
\rho_{{\mathrm{space}}}(u)=\int_{T}\rho(u,t)\,
\mathrm{d}t,\quad \rho_{{\mathrm{time}}}(t)=\int_{W}\rho(u,t)\,
\mathrm{d}u.
\end{align}
Using  the above marginal intensities, the conditional spatial and temporal intensities for any given time $t$ and for any given spatial location $u$, are respectively defined by 
\begin{align}
\label{eq:et:cond}
\rho(u|t)=\rho(u,t)/\rho_{{\mathrm{time}}}(t),
\quad \rho(t|u)=\rho(u,t)/\rho_{{\mathrm{space}}}(u).
\end{align}
Nonparametric kernel estimates of $\rho_{{\mathrm{space}}}$ and $\rho_{{\mathrm{time}}}$ are respectively given by
\begin{align}\label{eq:rhohat_space}
\hat\rho_{space}(u)=\sum_{i=1}^nk^2_{\epsilon}(u-u_i)/{C_{W,\epsilon}(u_i)}
\end{align}
and
\begin{align}\label{eq:rhohat_time}
\hat\rho_{time}(t)=\sum_{i=1}^nk^1_{\delta}(t-t_i)/{C_{T,\delta}(t_i)},
\end{align}
where $k_b^d$ is a $d$-dimensional kernel with bandwidth $b>0$, i.e.\ $k_b(v)=k(v/b)/b^d$
where $k$ is a given  density function, and 
$ C_{W,\epsilon}(u_i)=\int_W k^2_{\epsilon}(u-u_i)\de u$ and $ C_{T,\delta}(t_i)=\int_T k^1_{\delta}(t-t_i)\de t$ are edge correction factors in space and time, respectively (see details e.g.\ in \citep[page 168]{Baddeley:etal:16}).
In general, 
a nonparametric kernel estimate of the non-separable spatio-temporal intensity function
is 
\begin{align}\label{eq:rhohat}
\hat\rho(u,t)&=\sum_{i=1}^n\frac{ k^2_{\epsilon}(u-u_i)}{C_{W,\epsilon}(u_i)}\frac{k^1_{\delta}(t-t_i)}{C_{T,\delta}(t_i)}.
\end{align}
The kernels $k^1$ and $k^2$ of the above formulas may have different forms. However, in this work, we 
use isotropic Gaussian kernels. 
\section
{First-order spatio-temporal separability} \label{sec:first-order}
Recall that the point process $X$ has intensity function $\rho(u,t)$ for $(u,t)\in\mathbb{R}^2\times\mathbb{R}$. Hence, the intensity measure $\mu$ for $X$ can be written as 
\begin{align}\label{eq:intsMeas}
  \mu(A\times B)= E(n(X\cap(A\times B)) =\int _{A\times B} \rho(u,t)\de u \de t, \quad (A\times B)\subseteq \mathbb{R}^2\times \mathbb{R},
\end{align}
where $n(A)$ denotes the number of points of $X$ in any bounded set $A\subseteq \mathbb{R}^2\times \mathbb{R}$.
Note that, for a homogeneous 
STPP $X$, the intensity function $\rho$ is the mean number of points per unit space-time volume.
It is usually
assumed that the spatio-temporal intensity function of a STPP is separable, i.e.,
\begin{equation}\label{eq:first}
\rho(u,t)=\rho_1(u)\rho_2(t),\quad
\mbox{$(u,t)\in \mathbb{R}^2\times\mathbb{R}$,}
\end{equation}
 where $\rho_1$
and $\rho_2$ are non-negative  measurable functions. Under this assumption, for Borel sets
$A \subseteq \mathbb{R}^2$ and $B \subseteq \mathbb{R}$, the intensity measure \eqref{eq:intsMeas}  is a
product measure, that is
$\mu(A\times B)= \int_A\rho_1(u)\,\mathrm du\,\int_B\rho_2(t)\,\mathrm dt$   \citep[see more details in][]{moeller:ghorbani:12}. Some literature use the term {\em first-order spatio-temporal separability} instead of  {\em separability of the spatio-temporal intensity function}. In this paper we use the short term first-order separability.

The  null hypothesis of the first-order separability  implies that the intensity functions of the spatial and temporal component
processes in \eqref{eq:et:compnts} can be written as 
\begin{equation}\label{e:margins}
\rho_{{\mathrm{space}}}(u)=\rho_1(u)\int_{T}\rho_2(t)\,
\mathrm{d}t,\quad \rho_{{\mathrm{time}}}(t)=\rho_2(t)\int_{W}\rho_1(u)\,
\mathrm{d}u,
\end{equation}
and then by combining \eqref{eq:first} and \eqref{e:margins}, the intensity function of the point process $X$ under the first-order separability is
\begin{align}\label{eq:rhoSep}
 \rho_{sep}(u,t)=
\frac{\rho_{{\mathrm{space}}}(u)\rho_{{\mathrm{time}}}(t)}{\int_{W\times T
}\rho(u,t)\, \mathrm{d}(u,t)}.
\end{align}
Combining \eqref{eq:intsMeas} and \eqref{eq:rhoSep}, and defining 
$\mu_{\rm space}(A)=\int_A\rho_{\rm space}(u)\de u$ and $\mu_{\rm time}(B)=\int_B\rho_{\rm time}(t)\de t$, the intensity measure of $X$ under the first-order separability denoted by $\mu_{sep}$ is given by
\begin{align}\label{eq:intsMeasSep}
  \mu_{sep}(A\times B)= \frac{\mu_{{\mathrm{space}}}(A)\mu_{{\mathrm{time}}}(B)}{\mu(W\times T)}.
\end{align}
We further assume that the expected number of observed points is a positive and finite number, i.e., $0<\int_{W\times T}\rho(u,t)\, \mathrm{d}(u,t)<\infty$.
Furthermore, for a first-order separable model, the conditional spatial and temporal intensities in \eqref{eq:et:cond} are simplified to 
\begin{align*}
\rho(u|t)=
\frac{\rho_{\mathrm{space}}(u)}{\int_\mathrm{W}\rho_{\mathrm{space}}(u)\mathrm{du}},
\quad \rho(t|u)=
\frac{\rho_{\mathrm{time}}(t)}{\int_\mathrm{T}\rho_{\mathrm{time}}(t)\mathrm{dt}},
\end{align*}
which do not vary by changing time and space, respectively. 

In the spatio-temporal point processes context, it is usually assumed that the ``first-order spatio-temporal separability is a convenient working hypothesis which is hard to check" \citep{Gabriel:Diggle:09,moeller:ghorbani:12}.
Here, we propose formal tests for the first-order separability hypothesis. The test statistics are based on the separable and non-separable intensity estimates.
\section{Permutation tests for first-order separability}\label{sec:permutation-test}
\subsection{Test functions }\label{sec:first-test}
To test the first-order separability  hypothesis, we propose the following test function:
\begin{align}
\label{eq:Sfun}
 S(u,t)=\frac{\hat\rho(u,t)}{ \hat\rho_1(u)\hat\rho_2(t)}
=\frac{\hat\rho(u,t) } {\hat{\rho}_{{\mathrm{space}}}(u)\hat{\rho}_{{\mathrm{time}}}(t)/n},\quad  (u,t)\in W\times T,
\end{align}
for $\hat\rho(u,t), \hat{\rho}_{{\mathrm{space}}}(u), \hat{\rho}_{{\mathrm{time}}}(t) > 0$. The important part of this test function is the non-separable intensity in the numerator, while the separable intensity in the denominator is employed only for the scaling purpose. 
Because under the null hypothesis, $$\hat\rho(u,t)=\hat\rho_{sep}(u,t)=\frac{\hat\rho_{{\mathrm{space}}}(u)\hat\rho_{{\mathrm{time}}}(t)}{n},$$ thereupon, for all $(u,t)\in W\times T$, 
$S(u,t)=1$. Therefore, values deviating from one relate to non-separable structures. In particular, $S(u,t)>1$ indicates increased estimated non-separable intensity at the specific location $u$ and time $t$ in comparison to the estimated separable intensity.

The function $S(u,t)$ is a three-dimensional function which contains all the information about the intensity of the point process under study. In practice  the intensities and thus $S(u,t)$ are estimated on a finite number of spatial locations $u\in W$ and times $t\in T$. When the temporal region $T$ is small or low resolution in $T$ is adequate, it is convenient to visualize the three-dimensional function by plotting $S(u,t)$, $u\in W$, for each discretized time point $t$ (see Section~\ref{sec:performance} for details).
However, when $T$ is large or fine discretization is desired, 
instead of studying a large number of images, it can be of interest to consider the following one- and two-dimensional functions
that integrate $S(u,t)$ with respect to $u$ and $t$, namely
\begin{align}\label{eq:genTestStat_t}
   S_{\mathrm{time}}(t)&=\int_W S(u,t) {\rm d}u
\end{align}
and
\begin{align}\label{eq:genTestStat_u}
   S_{\mathrm{space}}(u)=\int_T S(u,t) {\rm d}t,
\end{align}
which can be easily visualized. On the other hand, some detailed information may be lost in the integration, as will be discussed in Section \ref{sec:performance}.
Under the hypothesis of first-order separability,
$S_{\mathrm{time}}(t)=|W|$  for all $t\in T $ and $S_{\mathrm{space}}(u)=|T|$ for all $u\in W $.
\subsection{Simulations under the first-order separability}\label{se:permut}
Because the distributions of the above test functions are not known, we need to resort to simulation based tests as usual in spatial and spatio-temporal statistics.
A test of first-order separability can be implemented  using permutations under the assumption of the inhomogeneous Poisson model.
The permutation test does not require distributional assumptions, or any type of model specification for the separability structure. 
The permutation test relies on the fact that if the null hypothesis is true, then shuffled data sets should look like the observed data, otherwise they should look different. In other words, the joint distribution of the observed data must be the same as the joint distribution  of the permuted data for any permutation, which is called permutation invariance. Following \cite{diggle:13}
we assume that the process is an inhomogeneous spatio-temporal Poisson process. Under this working assumption, it is easy to show that permutation invariance holds. To justify this claim, consider the density of the inhomogeneous spatio-temporal Poisson process with respect to the unit rate Poisson process as given by 
$
f(x)=e^{|\s|-\int_{\s}\rho(u,t)\de u\de t}\underset{{(u,t)\in x}}{\prod}\rho(u,t)
$
for finite point configurations $x \subset \s$.
Under the hypothesis of the first-order separability \eqref{eq:first},
\begin{align}
    f(x)=e^{|\s|-\int_{\s}\rho_1(u)\rho_2(t)\de u\de t}\prod_{(u,t)\in x}\rho_1(u)\rho_2(t).
    \label{eq:dens.sep}
\end{align}
Clearly, the value of the joint density function $f(x)$ given in \eqref{eq:dens.sep} for any random permutations of  the temporal points $t_i$ holding the locations $x_i$ fixed will not change.  The same holds for random permutations of  the $x_i$ holding the $t_i$ fixed.
Hence, simulations under the first-order separability hypothesis can be obtained using either random permutations of $x_i$ holding $t_i$ fixed or random permutations of $t_i$ holding $x_i$ fixed, with equal probability for each possible permutation. In all our tests, we permuted $t_i$s holding $x_i$s fixed.
If the null hypothesis of first-order separability is true, then
these randomly permuted data sets are statistically equivalent to the original data, and the rationale of the Monte Carlo test applies \citep{Barnard:63,Besag:Diggle:77,Baddeley:etal:16}.

Note that, in practice we observe only one realization of $X$, so it is not possible to distinguish a Cox process from its corresponding inhomogeneous Poisson process and thus considering inhomogeneous Poisson process as a working assumption seems reasonable. 
\subsection{Global envelope tests}\label{sec:GET}
In the Monte Carlo test, first a test statistic must be chosen and be calculated for the data and for $n$ simulations generated under the null hypothesis. Thereafter, the extremeness of the data statistic among all the statistic must be determined. This is done by ordering the statistics by some measure. We used the global envelope tests  \citep{mari:etal:13, mrkvicka:etal:2018, MyllymakiMrkvicka2019} to perform Monte Carlo tests based on $S(u,t)$, $S_{\mathrm{space}}(u)$ and $S_{\mathrm{time}}(t)$. An advantage of the global envelope tests is that they provide graphical interpretation of the test results. Furthermore, their non-parametric nature ensures that they are not sensitive to the inhomogeneity in the distribution of the test functions over the spatial, temporal or spatio-temporal domain.

In practice, the global envelope tests use the discretized version of the test function, say $S$. Let us denote the discretized test function calculated from the data by $S_1$ and the test functions calculated from the $n$ simulations by $S_2,\ldots,S_{n+1}$. 
In this paper, each of the $S_i$  can be thought as a $d$-variate vector, $S_i=(S_{i1},\dots,S_{id})$, where $d$ is given by the number of the grid points and it determines the  argument values $(u_1,t_1),\dots,(u_{d}, t_{d})$, $u_1,\dots,u_{d}$ or $t_1,\dots,t_{d}$ at which the function $S(u,t)$, $S_{\mathrm{space}}(u)$ or $S_{\mathrm{time}}(t)$ are evaluated, respectively.

We used the global extreme rank length (ERL) envelope test \citep{mrkvicka:etal:2018, MyllymakiMrkvicka2019}.
The test is based on the corresponding ERL measure suggested independently by \citet{mari:etal:13} and \citet{narisetty:nair:2016}: 
Let $(R_{i1}, R_{i2}, \dots , R_{id})$ stand for the pointwise ranks of $S_i$, and $\mathbf{R}_i=(R_{i[1]}, R_{i[2]}, \dots , R_{i[d]})$ be the same pointwise ranks ordered from smallest to largest, i.e.\ $R_{i[j]} \leq R_{i[j^\prime]}$ whenever $j \leq j^\prime$. 
The ERL ordering corresponds to the lexicographic ordering of $\mathbf{R}_i=(R_{i[1]}, R_{i[2]}, \dots , R_{i[d]})$. In general, for two vectors $v, v'\in \R^d$ lexicographic order is defined as $(v\prec v')$  if $\exists\, i\in\{1,2,\ldots,d\}, (\forall j<i\, v_j=v'_j)\wedge (v_i<v'_i)$. Specifically, an ERL measure
\begin{equation*}
   M_i = \frac{1}{n+1}\sum_{i^\prime=1}^{n+1} \mathbf{1}(\mathbf{R}_{i^\prime} \prec \mathbf{R}_i),
\end{equation*}
where $n+1$ plays only the role of a scaling factor,
can be attached to $S_i$, where a small value indicates extremeness of $S_i$. Here and in what follows ${\mathbf 1}(\cdot)$ denotes the indicator function.

The global envelope is constructed as a hull of those vectors $S_i$ that are considered non-extreme by the ERL measure $M_i$.
Formally, let $m_\alpha \in \mathbb{R}$ be the largest of the measures $M_i$ calculated for $S_1,\dots,S_{n+1}$ such that the number of those $i$ for which $M_i< m_{(\alpha)}$ is less or equal to $\alpha s$, and
let $I_\alpha = \{i\in 1,\dots, s: M_i \geq m_{(\alpha)} \}$ be the index set of vectors less or as extreme as $m_\alpha$. Then the $100(1-\alpha)$\% global envelope induced by the $M_i$ is given by
\begin{align*}
S^{(\alpha)}_{\lo, \, k}= \underset{i \in I_\alpha}{{\min}}\ S_{i k}
\quad\text{and}\quad
S^{(\alpha)}_{\up, \, k}= \underset{i \in I_\alpha}{{\max}}\ S_{i k} \quad \text{for } k = 1, \ldots , d.
\end{align*}
If the data vector $S_1$ does outside the envelope, then also $M_i < m_\alpha$ \citep[see e.g.][]{MyllymakiMrkvicka2019}. The Monte Carlo $p$-value can be calculated as $p =  \sum_{i=1}^{n+1} {\mathbf 1} (M_i \leq M_1)  \big/ (n+1)$.

We included in our simulation study also the global envelopes based on the Area measure suggested by \citet{mrkvicka:etal:2019}. However, in our examples the ERL and Area measures led to rather similar outcomes, and thus we decided to include the results based on the ERL measure only.

\subsection{Deviation test}\label{sec:deviationtests}
In addition to the global envelope tests, we included in our simulation study 
the deviation test based on the integral deviation measure
\begin{align}\label{eq:genTestStat_Sd}
   S_{d}=\int_{ W\times T } \left| \hat\rho(u,t)-\hat\rho_{sep}(u,t) \right| {\rm d} u {\rm d}t, 
\end{align}
which summarizes the discrepancy between the estimate of the spatio-temporal intensity function and its counterpart under the separability hypothesis.
This measure stems from testing the independence of two continuous random variables \citep[][p.\ 451]{blum:etal:61, Diggle:Gabriel:10}.
In practise, the integral in \eqref{eq:genTestStat_Sd} is replaced by a sum. 

A standard Monte Carlo test can be based on the statistic $S_{d,1}$ calculated from data and its simulated counterparts $S_{d,2},\dots,S_{d,n+1}$. Given that large values of $S_d$ are significant, the $p$-value of the test is $p =  \sum_{i=1}^{n+1} {\mathbf 1} (S_{d,i} \geq S_{d,1})  \big/ (n+1)$.
\section{$\chi^2$-test for first-order separability}\label{sec:chisquare}
A test of the first-order separability hypothesis can be performed without permutations in a manner similar to independence test in two-way contingency tables. The test is conditional on the observed number of points and asymptotically exact for Poisson processes.
In this spirit it is closely related to the well-known $\chi^2$-test of Complete Spatial Randomness.

We consider a division of the interval $T$ into disjoint sub-intervals $T_1, \ldots, T_J$ and similarly a division of the window $W$ into disjoint subsets $W_1, \ldots, W_I$. Both divisions can be based, e.g., on quantiles of the temporal and spatial coordinates, respectively, but are in fact arbitrary. Let $n_{ij}$ denote the number of observed events in the cell $W_i \times T_j$. Under the first-order separability assumption and conditionally on the observed number of events being $n$, we observe a binomial point process on $W \times T$ with probability density function proportional to 
$\rho_1(u) \rho_2 (t)$. In this setting the events are independent and the expected count in the cell $W_i \times T_j$ is
\begin{align*}
    e_{ij} = \frac{\left(\sum_{i=1}^I n_{ij}\right)\left(\sum_{j=1}^J n_{ij}\right)}{n}
\end{align*}
for each $i=1, \ldots, I, j = 1, \ldots, J$. Hence the test statistic
\begin{align*}
    \chi^2 = \sum_{i=1}^I \sum_{j=1}^J \frac{(n_{ij} - e_{ij})^2}{e_{ij}}
\end{align*}
has asymptotically, with the increasing   number of observations in the fixed observation window $W \times T$,
the $\chi^2$ distribution with $(I-1)(J-1)$ degrees of freedom. Recall that the test requires large sample sizes to be accurate. A simple rule of thumb regarding sample size is that the expected cell counts should be at least five. After defining the cells $W_i \times T_j$ and determining the observed counts $n_{ij}$ it is possible to perform the test using standard statistical software. We here utilized the R package {\em spatstat} \citep{Baddeley:etal:16}.

\section{Stochastic reconstruction for testing  first-order separability}\label{sec:SR}
We have also investigated other approaches for obtaining Monte Carlo replications than the permutation strategy described above. Among others, we explored various versions of the classical random shift approach \citep{Lotwick:Silverman:82,MrkvickaEtAl2020}, but none of these was able to produce replications \emph{under the null hypothesis}, i.e.\ with separable first-order structure, while preserving the interaction structure of the observed pattern.
However, the stochastic reconstruction procedure can be used to produce independent replications (outputs) with the same interaction structure as the observed data (input) and the same intensity functions of the spatial and temporal component process $X_{{\mathrm{space}}}$ and $X_{{\mathrm{time}}}$, respectively \citep{TscheschelStoyan2006,WiegandEtAl2013,KonasovaDvorak}.

The user chooses a set of summary characteristics that should be preserved by the reconstruction procedure. We suggest to use the square root of a non-parametric estimator of  
the inhomogeneous space-time $K$-function \citep{Gabriel:Diggle:09}, together with the separable estimator 
of the intensity function \eqref{eq:rhoSep}. Based on the experience from \citet{KonasovaDvorak}, we also suggest to use as further summaries the values $\hat{D}_k(r,t)$ giving the fraction of observed points which have at least $k$ neighbours within distance $r$ (in the spatial domain) and within the lag $t$ (in the temporal domain). These are considered only to be empirical characteristics describing interpoint distances rather than being estimators of some theoretical quantities. However, they are closely related to the raw estimates of the $k$th nearest neighbour distribution functions in a stationary space-time point process.

A so-called \emph{energy functional}, quantifying the dissimilarity between the input pattern $X$ and another pattern $Y$, is then constructed:
\begin{align}
E(X , Y ) & = w_{K} \int_0^{T_K} \int_0^{R_K} \left[ \sqrt{\widehat{K}(X;r,t)} - \sqrt{\widehat{K}(Y;r,t)} \right]^2 \, \mathrm{d} r \, \mathrm{d} t \label{eq:energy} \\
& + \sum_{k=1}^{k_{max}} w_{D_k}\int_0^{T_D} \int_0^{R_D} \left[ \widehat{D}_k(X;r,t) - D_k(Y;r,t) \right]^2 \, \mathrm{d} r \, \mathrm{d} t \nonumber \\ 
& + w_\Delta \sum_{i=1}^I a \left[ \hat{\rho}_{sep} \left( X; u_i, t_i\right) - \hat{\rho}_{sep} \left( Y; u_i,t_i\right)\right]^2, \nonumber
\end{align}
where $w_K, w_{D_k}, w_\Delta$ are the weights determining the relative importance of the individual terms, $\{ (u_1,t_1), \ldots, (u_I,t_I) \}$ are the center locations of the cells of a regular grid covering $W \times T$, $a$ is the volume of the grid cell and $T_K, R_K, T_D, R_D$ and $k_{max}$ are user-selected tuning constants.

The procedure starts with a binomial pattern $Y_0$ generated as a collection of $n$ independent points (the same as the number of observed points in pattern $X$) following a probability density function proportional to $\hat{\rho}_{sep}(X; u,t)$. Then iteration steps are repeated in which a new pattern  $Y^{new}$ is proposed by randomly deleting one point from the current pattern, say $Y_m$, and generating a new point in $W\times T$ with density again proportional to $\hat{\rho}_{sep}(X; u,t)$. 
The proposal is accepted if $E\left(X , Y^{new}\right) \leq E\left(X , Y_m \right)$, otherwise it is rejected. The algorithm stops when a user specified stopping rule is met, e.g.\ 
after performing a maximum allowed number of iterations or after rejecting a certain amount of proposals in a row \citep{TscheschelStoyan2006,illian:penttinen:stoyan:stoyan:08}. By minimization of the energy functional the output pattern $Y^{out}$ is forced to have approximately the same interaction structure as the input pattern $X$ (as described by the $K$- and $D_k$-functions) while having a separable first-order structure (as described by $\hat{\rho}_{sep}$).

After a large number of independent output patterns is generated, these can be used to perform a Monte-Carlo test of the first-order separability hypothesis as in Section~\ref{sec:permutation-test}. The outputs can be considered to be independent replicates of the data obtained under the null hypothesis. The performance of the Monte-Carlo test of course relies on the interaction structure of the observed data being correctly captured by the reconstruction procedure.

Using the stochastic reconstruction procedure requires some tuning of the parameters. It is also strongly suggested to verify on simulated data that the outputs of the reconstruction algorithm have the same properties as simulations from the correct model. This can be done following the suggestions of \cite{KonasovaDvorak} and will be illustrated in the data example below.
\section{Performance of separability tests}\label{sec:performance} 
\subsection{Separability of inhomogeneous Poisson processes}\label{sec:sepInhomP}
We investigated the performance of the global envelope tests based on the test functions \eqref{eq:Sfun}-\eqref{eq:genTestStat_u},
the deviation test $S_d$ based on \eqref{eq:genTestStat_Sd} as well as the $\chi^2$-test in the following simulation study:
We let $W\times T = [0,1]^2\times[0,1]$ be the observation window and considered the inhomogeneous spatio-temporal Poisson process within $W\times T$ with the following intensity:
\begin{align}
 \rho (u,t) = (\nu - \gamma)\xi(u)\psi(t) + \gamma  \phi_{\boldsymbol{\mu},\boldsymbol{\Sigma}}(u,t),
\label{eq:sim.rho1}
\end{align}
where the parameter $\nu$ controls the expected number of points in the pattern and the parameter $\gamma\in[0,\nu/2]$ controls the degree of separability. 
For $\gamma=0$, the model is separable as all the points come from the first part of the model, whereas $\gamma=\nu/2$ corresponds to the most non-separable model where half of the points come from the first and another half from the second part of the model. 
In our simulation study, we considered different values of $\gamma=0,25,50,75,\dots,200$ and we determined $\nu$ by fixing the expected number of points in the pattern to be around 600, to have a corresponding number of points as in our data example, in all simulations.
The rationale behind the model \eqref{eq:sim.rho1} is to mimic a situation in practical applications, e.g.\ in environmental epidemiology, where the incidents typically occur randomly in the population, possibly with some general trend in time, but at a certain time there occurs a sudden burst of incidents around a contaminated source or another starting location of an epidemic. 
One famous example of such situation is the cholera cases in the proximity of the polluted water pump in Soho, London 1854 \citep[][pages 118-122 and references therein]{Bivand:etal:13}. Another example is the UK 2001 foot and mouth disease data analysed in Section~\ref{sec:fmd}. In the first part of the model, $\xi(u)$ can be understood as a population density in space. Given that the points represent incidents of some contagious disease, in the case $\gamma=0$, they occur randomly in the population. On the other hand, $\psi(t)$ is the baseline trend in time.
The second part of the model with $\phi_{\boldsymbol{\mu},\boldsymbol{\Sigma}}(u,t)$ creates a ``burst'' of incidents around a particular location and time. 

More precisely, for the baseline spatial and temporal densities $\xi(u)$ and $\psi(t)$, we considered four different models: 
\begin{enumerate}
\item overall constant density with $\xi(u)=1$ and $\psi(t)=1$,
\item $\xi(u) = 1$ and the temporal density $\psi(t)$ was the univariate normal distribution $\phi_{\mu,\sigma}(t)$ with mean $\mu=0.5$ and variance $\sigma=0.2$,
\item $\psi(t) = 1$ and the spatial density $\xi(u)$ was the bivariate normal distribution $\phi_{\boldsymbol{\mu},\boldsymbol{\Sigma}}(u)$ with mean $\boldsymbol{\mu}=(0.5,0.5)$ and with diagonal covariance matrix $\boldsymbol{\Sigma}$ with variances $0.2$ in both directions,
\item $\psi(t) = \phi_{\mu,\sigma}(t)$ and $\xi(u) = \phi_{\boldsymbol{\mu},\boldsymbol{\Sigma}}(u)$ as in (ii) and (iii), respectively. 
\end{enumerate}
Thus, in the case (i), the baseline points occur uniformly in space and time, generating a point pattern which is homogeneous in both space and time for $\gamma=0$. This can be understood as a case where prevalence of some disease is inspected on a relatively small region with approximately constant population density.
In the case (ii), the number of incidents increases up to the time 0.5 after which the density again decreases gradually. 
The generated point pattern is homogeneous in space when $\gamma=0$, but inhomogeneous in time.
In the case (iii), the incidents occur in time at a constant rate, whereas the population density represented by $\xi(u)$ is centred in the middle of $W$, and thus the inhomogeneity occurs in space.
Finally, the case (iv) is a combination of the cases (ii) and (iii) where the base line density is inhomogeneous both in space and time.

The second part of the model with $\phi_{\boldsymbol{\mu},\boldsymbol{\Sigma}}(u,t)$ creates non-separability in the model when $\gamma\neq 0$.
The density $\phi_{\boldsymbol{\mu},\boldsymbol{\Sigma}}(u,t)$ is the three dimensional normal distribution with mean $\boldsymbol{\mu}= (0.3,0.3,0.2)$ and with diagonal covariance matrix $\boldsymbol{\Sigma}$ 
creating a ``burst'' of points around $\boldsymbol{\mu}$. 
We fixed the variance matrix to $\boldsymbol{\Sigma}=\text{diag}(0.05,0.05,0.05)$.

We used the independent thinning method to generate 1000 realisations of each of the spatio-temporal inhomogenous Poisson processes described above. Briefly, considering the fact that for each model the intensity function  $\rho(u,t)$ in \eqref{eq:sim.rho1}
is bounded  above by a positive constant $\rho_{\text{max}}$ on the given observation window $W\times T=[0,1]^2\times[0,1]$, we first simulated a homogeneous Poisson process $X_{\text{max}}$ within $W \times T$  with intensity $\rho_{\text{max}}$, and second made an independent thinning of $X_{\text{max}}\cap W \times T $ where the retention probability of a point $(u,t)\in X_{\text{max}}\cap W \times T$ is given by $p(u,t)=\rho(u,t)/\rho_{\text{max}}$  \citep[see e.g.][for details]{daley:vere-jones:08}. 

Considering each simulated pattern as our data on its own turn, we tested the first-order separability of it by the different tests. To generate simulations under the null hypothesis, we used the permutation procedure explained in Section~\ref{se:permut}
and we fixed the number of permutations to $1999$ in each case.
We used equations \eqref{eq:rhohat_space}-\eqref{eq:rhohat} to estimate the space, time and space-time intensities.  
The choice of the bandwidth in space was made to minimise the mean-square error criterion defined by \cite{Diggle:85}  and obtained using the function \texttt{bw.diggle()} of the R package  {\em spatstat} \citep{Baddeley:etal:16}. To choose the bandwidth in time we used 
the function \texttt{bw.SJ()} of the R package {\em base} \citep{R2019} which is the  \cite{sheather:jones:91} method of bandwidth selection using pilot estimation of derivatives to minimise the mean integrated square error criterion.
For the example pattern of Figure \ref{figure:examp} of the model \eqref{eq:sim.rho1} with case (iii), the bandwidths in space and time were respectively $0.025$ and $0.037$. 
We made the ERL 
global envelope tests and the deviation test
based on the function $S(u,t)$ at a $25\times 25\times 20$ grid covering $W\times T$. Furthermore the global envelope tests were performed for the functions $S_{\text{space}}(r)$ and $S_{\text{time}}(r)$ in the corresponding space and time grids. 
In our calculations, we utilized the R library GET \citep{mari:etal:13, MyllymakiMrkvicka2019}.
For the $\chi^2$-test we used the $4\times 4\times 4$ grids based on the quantiles of the temporal and spatial coordinates ensuring that the amount of points was in average more than five points in each grid cell.
For each test, we calculated the number of rejections of the null hypothesis among the 1000 repetitions at the significance level 0.05. 

Table \ref{tab:table1}  shows the results of the simulation study.  
The case $\gamma=0$ corresponds to empirical significance levels, while for the other values of $\gamma$, powers of different tests to detect deviation from the null hypothesis are given. The mean number of points, $\bar{n}$, in the 1000 simulated point patterns are shown at the third column of  the table.
The following observations can be done:
\begin{itemize}
    \item[(a)] As expected due to the theoretical result \eqref{eq:dens.sep}, the empirical significance levels of all tests were close to the nominal level 0.05. (They should be between $0.037$ and $0.064$ with a probability of 0.95 given by the 2.5\% and 97.5\% quantiles of the binomial distribution with parameters 1000 and 0.05).
    \item[(b)] The power of the tests increased with increasing degree of non-separability.
    \item[(c)] The $S$-function based tests had extremely high power for all models. However, the tests based on $S_{\text{space}}$ had lower performance for models (ii) and (iv), which had an inhomogeneous base line temporal density, while the tests based on $S_{\text{time}}$ had lower performance for models (iii) and (iv), which had an inhomogeneous base line spatial density. Thus, the temporal or spatial inhomogeneity had an effect on the performance of the tests based on $S_{\text{space}}$ and $S_{\text{time}}$. 
    \item[(d)] The power of the deviation test \eqref{eq:genTestStat_Sd} and the $\chi^2$-test was also high, but tended to be slightly lower than the power of the global envelope tests for the critical values $\gamma=25, 50$. 
\end{itemize}
The result (c) was expected, since the $S$-function utilizes all the information of the intensity function, while the latter two summarize $S$ and can therefore loose some of the information as illustrated  by the example in Section \ref{sec:simexample}.

We performed the permutation tests also with the global area envelope tests \citep{mrkvicka:etal:2019, MyllymakiMrkvicka2019} (results now shown here). Its power tended to be slightly higher for $S$ than the power of the global ERL envelope tests (for $\gamma=25$), while it was the other way around for $S_{\text{space}}$.
\begin{table}[h!]
  \begin{center}
    \caption{
The proportion of rejections among $1000$ repetitions of the permutation and $\chi^2$-tests at the significance level $\alpha = 0.05$ for the four models specified by \eqref{eq:sim.rho1} (i)-(iv). The global envelope and deviation tests were performed with $1999$ permutations of times using a $25\times 25\times 20$ grid, and the $\chi^2$-test was based on a $4\times 4\times 4$ grid.
    }
     \label{tab:table1}
    \begin{tabular}{l|c|c|c||ccc|c|c} 
    
&&&& \multicolumn{3}{c|}{Global ERL envelope}  &{Deviation}&{$\chi^2$ test}\\
        $\mathrm{Model}$ & 
        $\gamma$ & $\mathrm{\Bar{n}}$ & $\nu$ & $S(u,t)$ &
        $S_{\mathrm{space}}(u) $& 
        $S_{\mathrm{time}}(t)$&
        $S_{d}$&
 
        \\
       \hline\hline
       \multirow{10}{*}{1}
       &0&599.4&600&0.066&0.054&0.050&0.052&0.048\\
       &25&599.4&600&0.869&0.733&0.633&0.549&0.395\\
       &50&600.3&600&1.000&0.998&0.999& 0.994&0.864\\
       &75&600.1&600&1.000&1.000&1.000&1.000&0.984  \\
       &100&598.6&600&1.000&1.000&1.000&1.000 &0.996\\
       &125&601.3&600&1.000&1.000&1.000&1.000&1.000\\ 
       &150&600.5&600&1.000&1.000&1.000&1.000&1.000\\
       &200&599.4&600&1.000&1.000&1.000&1.000&1.000\\

\hline
       \multirow{10}{*}{2}
       &0 &599.8 &608&0.048&0.056&0.046&0.047&0.046\\
       &25&599.8 &608&0.965&0.077&0.835&0.774&0.786\\
       &50&600.3 &607&1.000&0.119&1.000&1.000&1.000\\
       &75&600.1 &607&1.000&0.147&1.000&1.000&1.000\\
       &100&601.5&607&1.000&0.121&1.000&1.000&1.000\\ 
       &125&599.4&606&1.000&0.250&1.000&1.000&1.000\\
       &150&600.5&606&1.000&0.720&1.000&1.000&1.000\\
       &200&600.4&606&1.000&0.999&1.000&1.000&1.000\\
\hline
       \multirow{10}{*}{3}
       &0&600.7&615&0.054&0.044&0.060&0.050&0.052\\
       &25&600.8&615&0.728&0.671&0.071&0.603&0.739\\
       &50&601.1&614&0.998&0.995&0.154&0.995&0.991\\
       &75&599.9&614&1.000&1.000&0.431&1.000&0.999\\
       &100&599.9&613&1.000&1.000&0.687&1.000&1.000\\
       &125&599.4&612&1.000&1.000&0.884&1.000&1.000\\
       &150&600.9&612&1.000&1.000&0.958&1.000&1.000\\
       &200&599.9&611&1.000&1.000&0.995&1.000& 1.000\\
\hline
       \multirow{10}{*}{4}
       &0&599.1&623&0.050&0.059&0.040&0.039& 0.054\\
       &25&600.9&623&0.900&0.106&0.060&0.757&0.487\\
       &50&600.0&622&0.999&0.111&0.269&0.999&0.982\\
       &75&601.1&621&1.000&0.118&0.637&1.000&1.000\\
       &100&600.5&620&1.000&0.085&0.859&1.000&1.000\\
       &125&600.8&619&1.000&0.221&0.965&1.000&1.000\\
       &150&599.8&618&1.000&0.617&0.993&1.000&1.000\\
       &200&602.1&617&1.000&0.992&1.000&1.000&1.000\\
    \end{tabular}
  \end{center}
\end{table}

\subsection{Graphical interpretation of the  first-order separability test results}\label{sec:simexample}

\begin{figure}[!htb]%
\begin{center}
\begin{tabular}{ccc}
\includegraphics[width=0.25\textwidth]{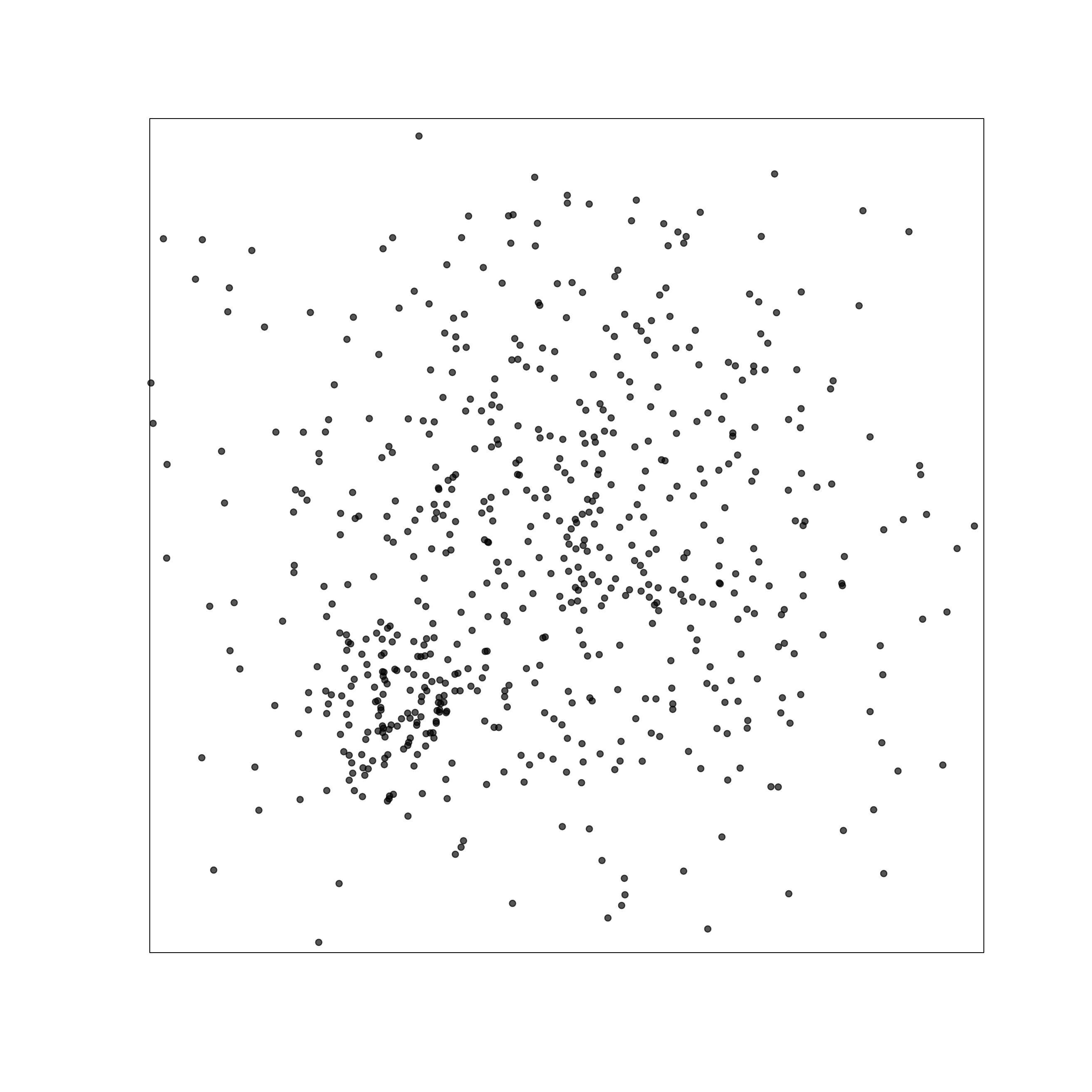}
\includegraphics[width=0.245\textwidth]{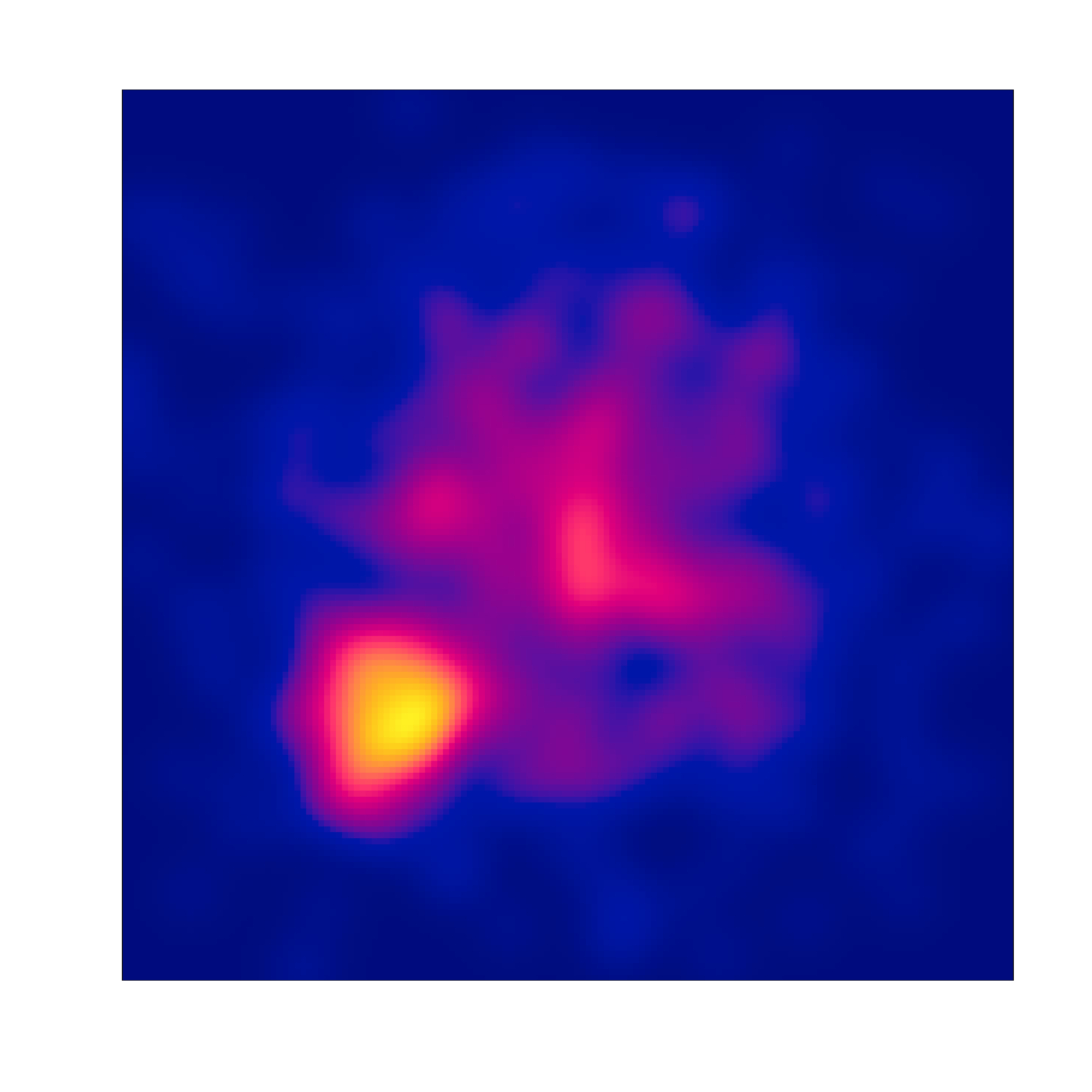}
\includegraphics[width=0.265\textwidth]{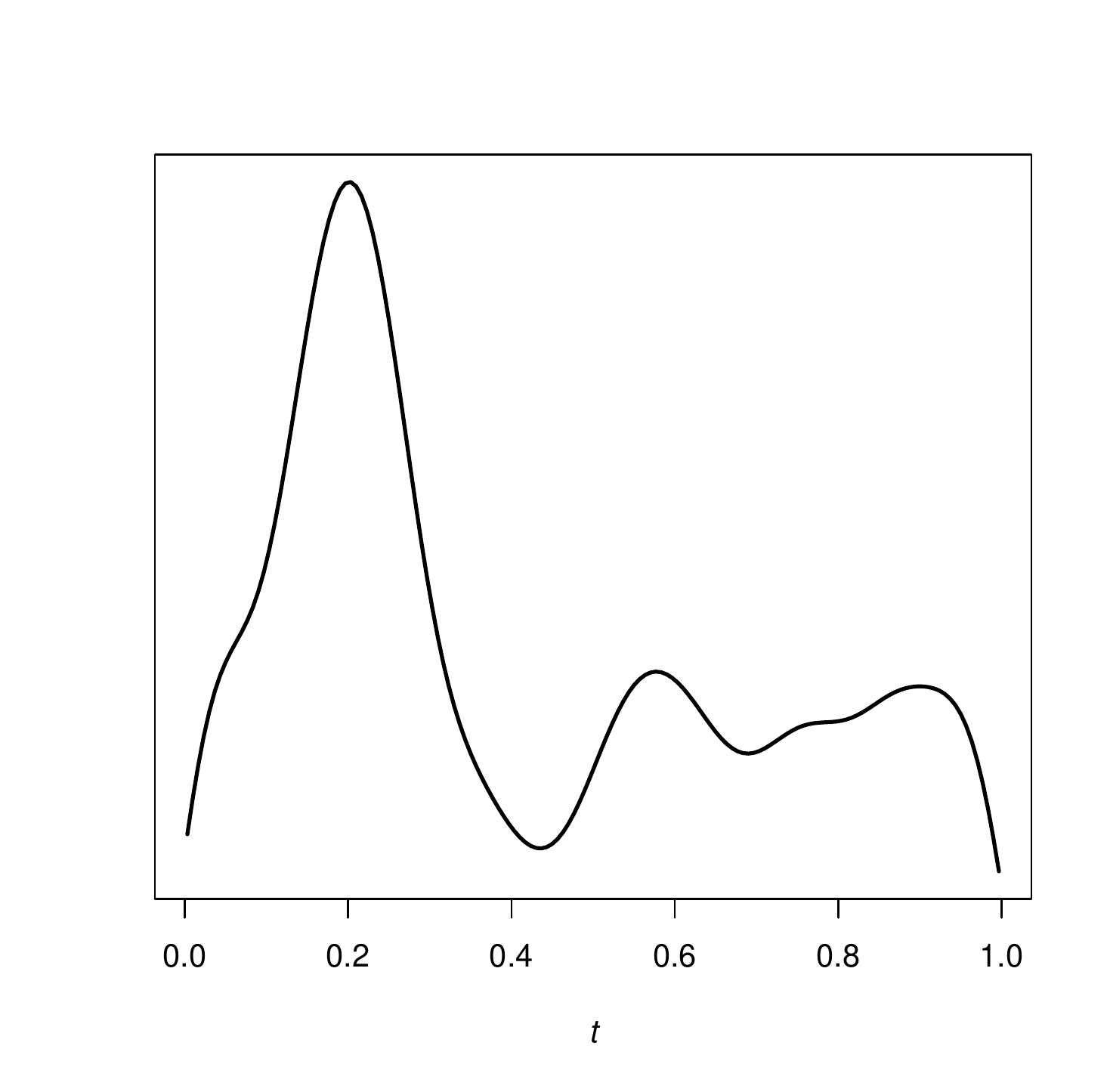}
 \end{tabular}
\caption{A realisation of the non-separable inhomogeneous Poisson process model \eqref{eq:sim.rho1} with the spatial and temporal base line densities of the case (iii) of Section \ref{sec:sepInhomP} and $\gamma=100$. The spatial component pattern (left), $\hat\rho_{\mathrm{space}}(r)$ (middle), and $\hat\rho_{\mathrm{time}}(t)$ (right).}\label{figure:examp}
\end{center}
\end{figure}

\begin{figure}[!htb]%
\begin{center}
\subfloat[Data]{\includegraphics[width=0.9\textwidth]{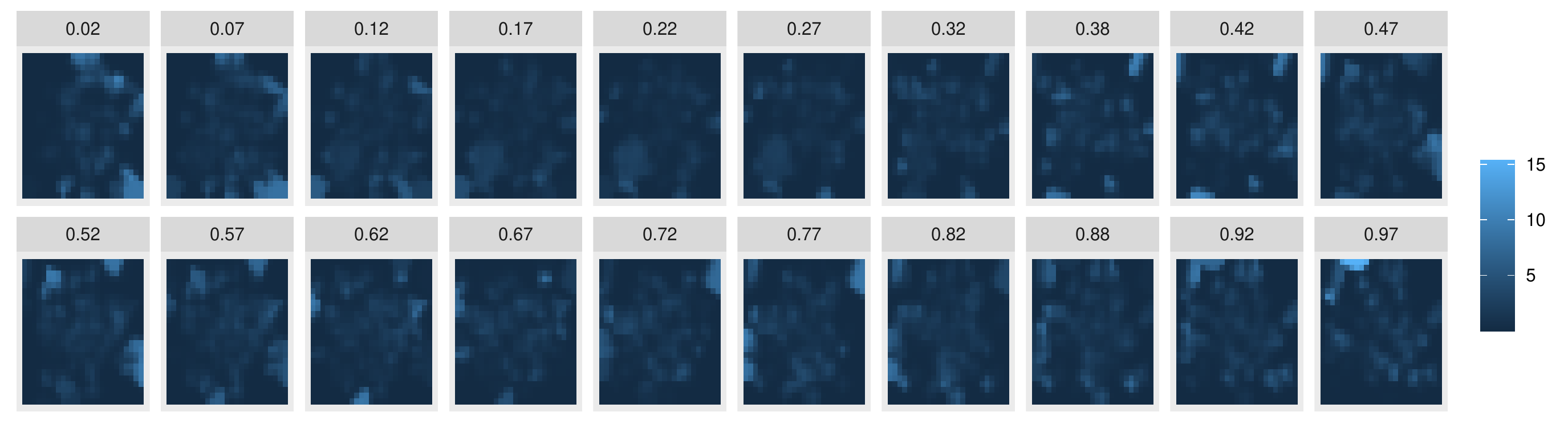}}\\
     \subfloat[Lower envelope]{\includegraphics[width=0.9\textwidth]{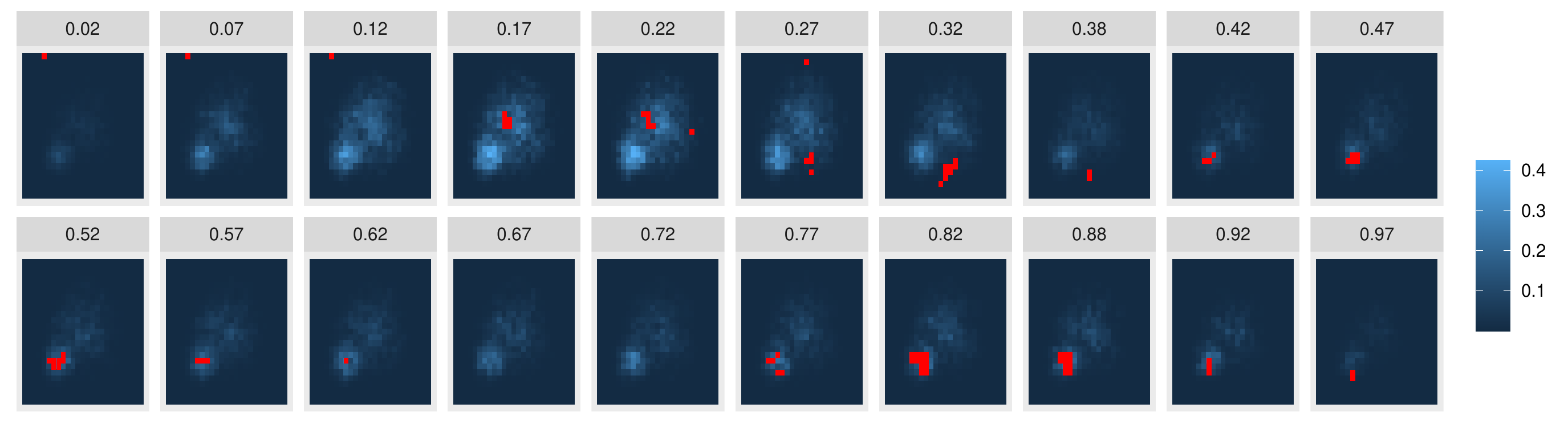}}\\
    \subfloat[Upper envelope]{\includegraphics[width=0.9\textwidth]{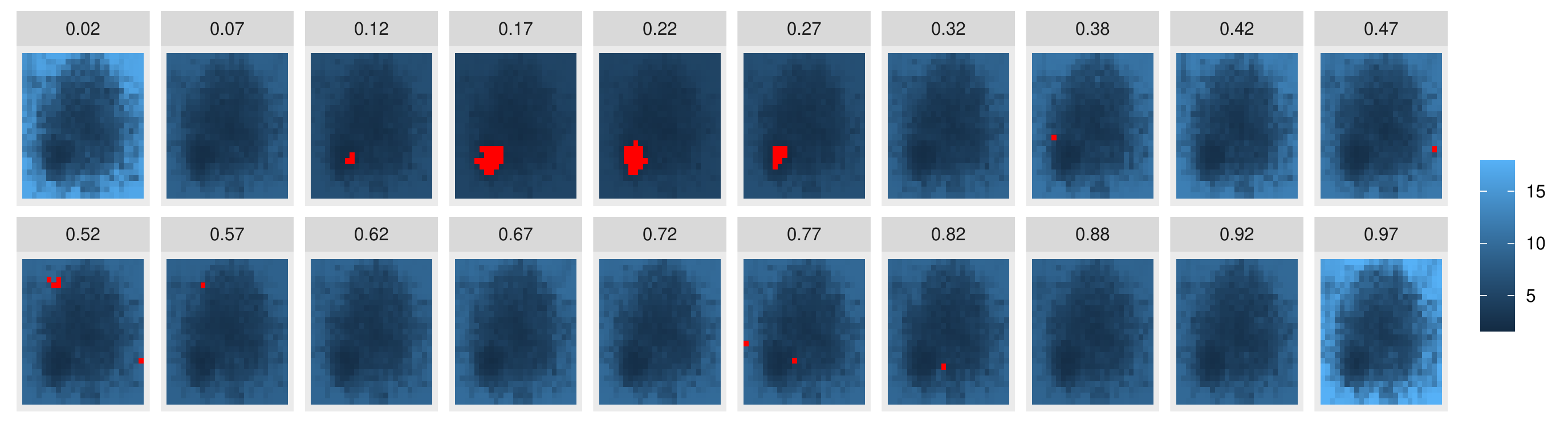}}
 \caption{Permutation based test: testing the first-order separability of  the point pattern of Figure \ref{figure:examp} using the global ERL envelope test with the function $S$ ($p=1\cdot 10^{-3}$): (a) The empirical $S$-function  (blue color), (b) the lower envelope overlaid by the significant regions (red) where the empirical function goes below the envelope, and (c) the upper envelope overlaid by the significant regions where the empirical functions goes above the envelope. The 95\% global envelope was constructed from 1999 simulations. The temporal coordinates of the grid are given above the corresponding plots.
}\label{fig:GET1}  %
\end{center}
\end{figure}

\begin{figure}[!htb]%
\begin{center}
\begin{tabular}{c}
\includegraphics[width=0.3\textwidth]{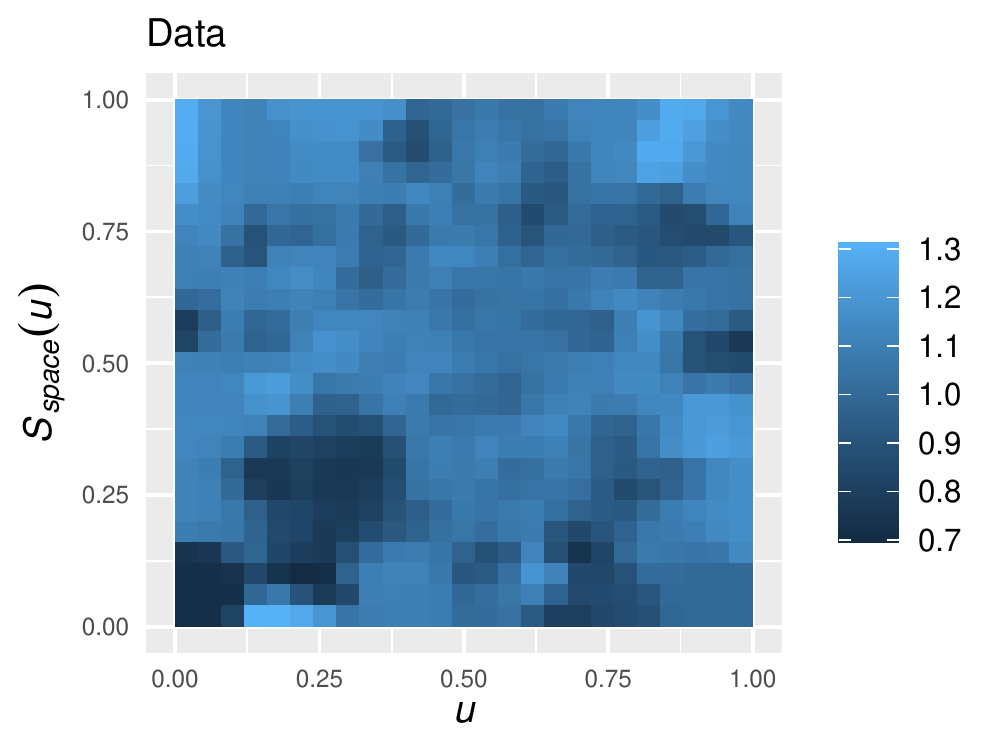}
\includegraphics[width=0.3\textwidth]{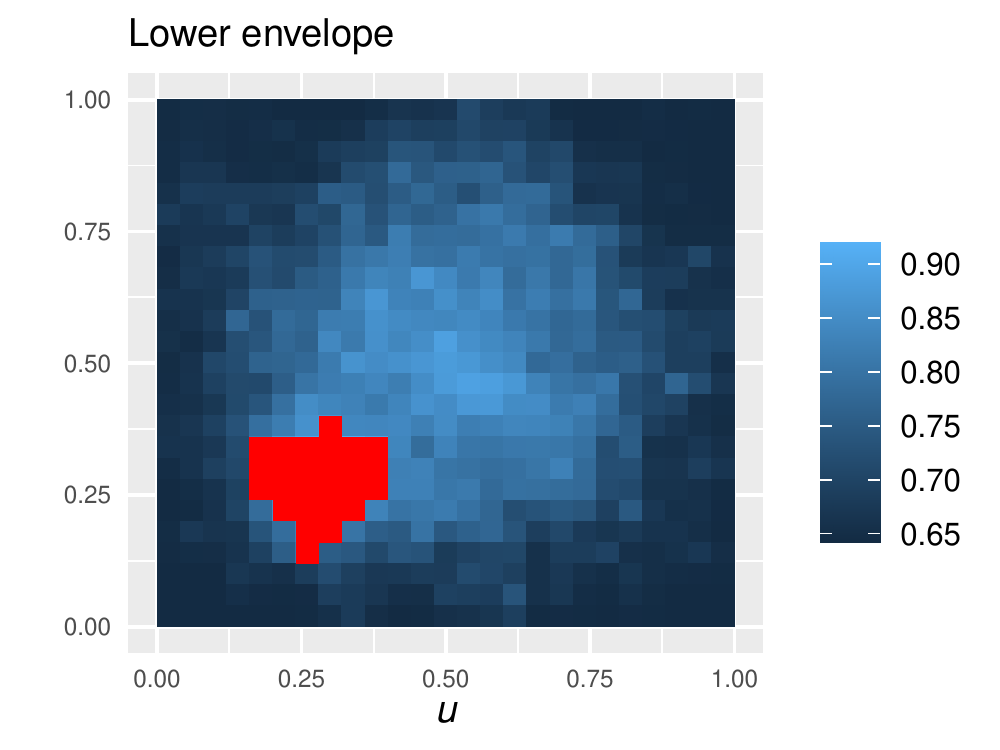}
\includegraphics[width=0.3\textwidth]{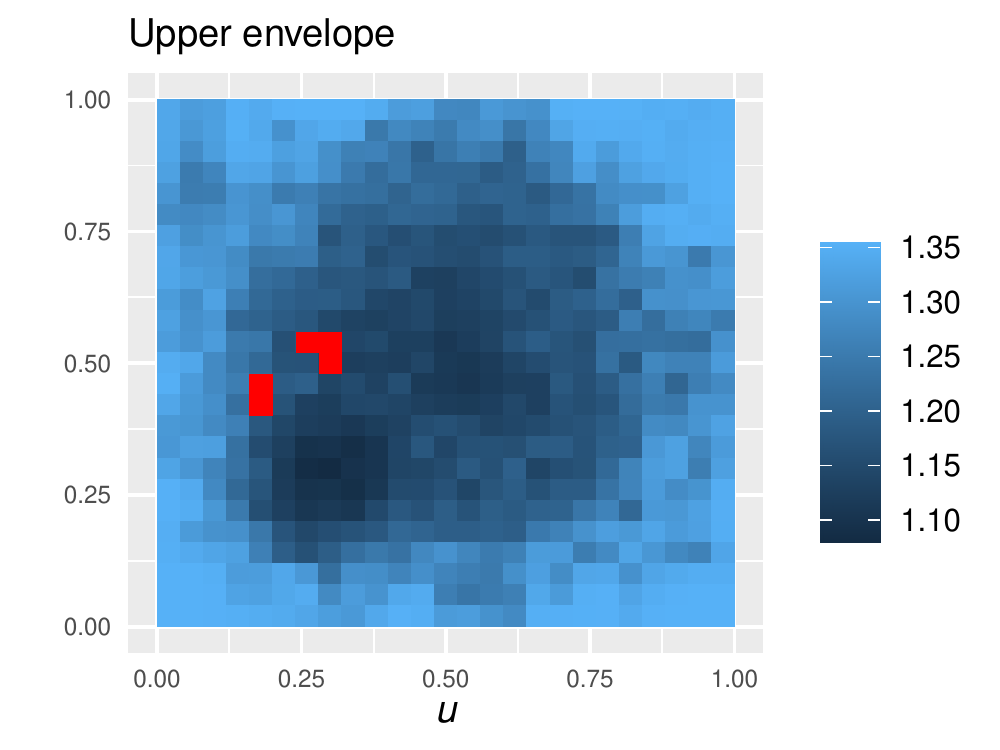}
 \end{tabular}
\caption{Permutation based test: testing the first-order separability of  the point pattern of Figure \ref{figure:examp} using the global ERL envelope test with the function $S_{\text{space}}(u)$ ($p=5\cdot 10^{-4}$):  The empirical $S_{\text{space}}$ function  (blue color)(left),  the lower envelope overlaid by the significant regions where the empirical function goes below the envelope(middle), and  the upper envelope overlaid by the significant regions where the empirical functions goes above the envelope(right). The 95\% global envelope was constructed from 1999 simulations.   }
  \label{fig:GET2}%
\end{center}
\end{figure}

\begin{figure}[!htb]%
\begin{center}
\begin{tabular}{c}
\includegraphics[width=0.5\textwidth, height=0.4\textwidth]{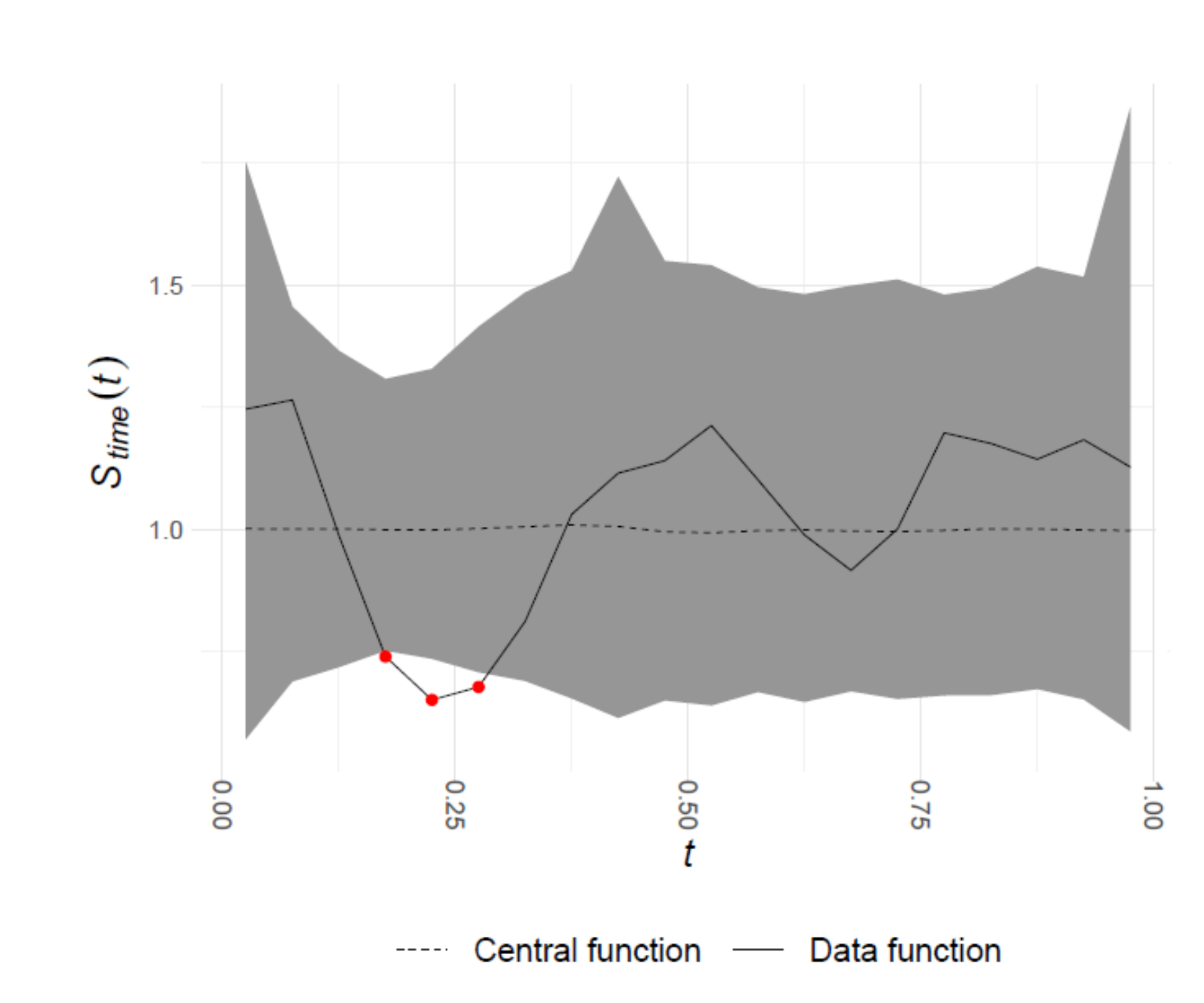}
 \end{tabular}
\caption{Test of first-order separability for the point pattern of Figure \ref{figure:examp} using the global ERL envelope test with the function $S_{\text{time}}$ ($p =25\cdot 10^{-4}$)  the observed function and the 95\% global envelope. The central function is the mean of all functions $S_{\text{time}}$.}
\label{fig:GET3}%
\end{center}
\end{figure}

Figure \ref{figure:examp} depicts a STPP simulated from the non-separable inhomogeneous Poisson process \eqref{eq:sim.rho1} with the spatial and temporal base line intensities (iii) 
and $\gamma=100$.
Figures \ref{fig:GET1}, \ref{fig:GET2} and \ref{fig:GET3} show the outputs of the three global ERL envelope tests based on the test functions $S$, $S_{\text{space}}$ and $S_{\text{time}}$, respectively. The test setup was as described above.

Figure \ref{fig:GET1} shows the regions where the $S$-function estimated from the data goes above and below the global envelope constructed from permutations. 
The result indicates increased intensity of points particularly around the spatial location $(0.3, 0.3)$ at times 0.12-0.27. On the other hand, there are several times where the data function goes below the envelope. The explanation for this test output is as follows: In the generated pattern, the intensity is highest in the burst region centred at $(0.3,0.3,0.2)$. The permuted patterns have lower intensity at this location and time, because the random permutations of times shuffle those points across the whole time window $T$. The shuffling simultaneously increases the intensity at the other times, whereby the data function goes below the envelope in the burst region at times different from the time of the burst. Because typically the interest lies in the increased numbers of points or incidents, the locations and times where the upper envelope is exceeded tend to be more interesting than exceeding the lower envelope. 
The few crossings of the upper envelope at later times have occurred just by a random chance; the choice of the rather small bandwidth ($0.03$) may have supported the detection of such small random clusters.

It should be noted that the separable intensity in \eqref{eq:Sfun} is the same for the data and all permutations and, therefore, it serves only as a scaling factor. Consequently, the exceeding of the envelope can be interpreted as increased intensity in comparison to the null hypothesis.

The function $S_{\text{space}}$ is able to detect the same spatial location around $(0.3,0.3)$ (see Figure \ref{fig:GET2}). At first thought, it might be surprising that the data function goes below the envelope, and not above. However, the function $S_{\text{space}}$ is obtained by integrating the $S$-function over all times and, as explained above, the $S$-function had different behavior at different times.
In this case, the function $S_{\text{time}}$ detected the deviation from the null hypothesis, as well: the data function went below the envelope at times 0.17, 0.22 and 0.27 (see Figure \ref{fig:GET3}).
Both  functions $S_{\text{space}}$ and $S_{\text{time}}$ will probably go below the envelope in many applications as well, when the non-separability occurs due to increased incidents at particular locations and times. We advice to use these functions only to detect the locations and times of non-separability and then draw further conclusions by inspecting the STPP.

An advantage of the non-parametric rank envelopes is that they adapt to the variability of the test function across its domain. This can be observed from the lower and upper envelopes that vary across the space or time domain (see Figures \ref{fig:GET1}, \ref{fig:GET2}, \ref{fig:GET3}).


\subsection{Separability for log Gaussian Cox processes}\label{sec:LGCP} 
Consider a spatio-temporal  stationary log Gaussian Cox process (LGCP) $X$ driven by  a non-negative stochastic process 
$$
\Lambda(u,t)= \exp(m(u,t) + Z(u,t)), \quad (u,t)\in\R^2\times\R,
$$
where 
$m(u,t)$ is a non-random/deterministic trend and $ Z(u,t)$ is a Gaussian random field with mean 0, variance $\sigma^2(u,t)$ and covariance function $C((u_1,t_1),(u_2,t_2))$ \citep{moeller:syversveen:waagepetersen:98}. For this process, the intensity and pair correlation functions are  respectively given by
\begin{align}\label{eq:LGCPrho}
\rho(u,t)=\exp(m(u,t)+\sigma^2(u,t)/2)
, \quad
g((u,s),(v,t))=\exp(C((u,s),(v,t))). 
\end{align}
We assume here that the LGCP $X$ is second-order intensity-reweighted stationarity (SOIRS), i.e.\ the pair correlation function $g$ depends only on spatial distance and time lag between two points \citep{baddeley:moeller:waagepetersen:97,Gabriel:Diggle:09}. 
This implies that the covariance function
$C((u,s),(v,t))=C(u-v,s-t)$
is stationary and the variance $\sigma^2(u,t) = C((u,t),(u,t)) = \sigma^2$ is constant. Under these assumptions, $\rho(u,t)$ is separable if 
the mean function $m(u,t)$ takes an additive structure, i.e. 
\begin{align}\label{eq:sepLGCP}
m(u,t)= \log\rho_1(u)+\log\rho_2(t)- \sigma^2/2=m_1(u)+m_2(t), 
\end{align}
where $m_1(u)$ and $m_2(t)$ are two functions for describing the trend of the field in space and time, respectively.
That is, if the first-order separability holds, then spatial and temporal covariates may be used for modelling trends in space and time, respectively, whereas otherwise one should look for spatio-temporal covariates.

We investigated the performance of the permutation based test and $\chi^2$-test for testing the first-order separability of a spatio-temporal LGCP on $W\times T = [0,1]^2\times[0,1]$. 
For the sake of simplicity, we specifically set
\begin{align}\label{eq:LGCP_m}
 m(u,t)=\beta_0+ \beta_1(x-t)+\gamma'xt,
\end{align}
where $\gamma'=0$ corresponds to the hypothesis of first-order separability, while for $\gamma'\neq 0$, the intensity function is non separable. We also set
\begin{align}\label{eq:Zut}
  Z(u,t) = \sigma_1Z_s(u)+\sigma_2Z_t(t)+\gamma'' Z_{st}(u,t),\qquad  u\in\R^2, t\in \R,
\end{align}
for parameters $\sigma_1,\sigma_2 > 0$, and $\gamma'' \ge 0$.
We further assumed that $Z_s$, $Z_t$, and $Z_{st}$ are independent Gaussian random fields  with mean zero
and covariance functions 
$C_1(u_1,u_2) =\exp (- \norm{u_1-u_2}^2/\phi_1)$, $C_2(t_1,t_2) = \exp(-|t_1-t_2|)/\phi_2)$, 
and $C_3((u_1,t_1),(u_2,t_2)) = C_1(u_1,u_2)C_2(t_1,t_2)$, respectively, with correlation parameters $\phi_1,\phi_2 > 0$.
It is worth mentioning that, setting $\gamma'=\gamma''=0$ implies that the realisation of the underlying random intensity $\Lambda(u,t)$ has multiplicative form. This in turn indicates that, conditional on $\Lambda(u,t)$, for $\gamma'=\gamma''=0$, a realisation of a first-order separable inhomogeneous Poisson process is obtained.
The special case of the above LGCP model assuming homogeneity ($m(u,t)=\log \rho$, with $\rho>0$ as a constant) has been used in \citet{Moeller:etal:19} to investigate the performance of the space-sphere $K$-function.
They fitted the LGCP model with $\gamma''=0$ and prepared a Monte Carlo test, the global rank envelope test, based on the simulations from the fitted model.
We explore here instead the performance of the completely non-parametric permutation based test and $\chi^2$-test.
 
First, we explored the empirical significance level of the tests, i.e.\ the performance of the tests for the case $\gamma'=0$. For each value of $\gamma''= 0,0.5,1$, we simulated $1000$ realisations of the LGCP on $[0,1]^2\times [0,1]$ with 
$\beta_1= 0.25$, $\sigma_1=\sigma_2=0.5$,
$\phi_1= 0.06$ and  $\phi_2= 0.05$.
Briefly, to simulate a realization of a LGCP with the given covariance function $C((u_1,t_1),(u_2,t_2))$ and mean 0, we first generated a realization of a Gaussian random field, $z(u,t)$, in a $20\times20\times 20$ grid with covariance function $C((u_1,t_1),(u_2,t_2))$ and mean 0 using the circulant embedding method \citep{wood:chan:94} in the R package {\em RandomFields} \citep{Schlather:etal:15}. Thereafter, we simulated a realisation of the inhomogeneous Poisson process with intensity $\exp(m(u,t) + z(u,t))$, $(u,t)\in\R^2\times\R$.
For each realisation, we tested the null hypothesis $\gamma'= 0$ (the first-order separability) using both the permutation based test and  $\chi^2$-test at the significance level 0.05. Note that the parameter values were chosen such that the mean number of the points of each realisation was around 200.
The permutation based test was based on the test function \eqref{eq:Sfun} with three different choices of the bandwidths for estimating the intensities, namely 
a) with the bandwidths selection of Section \ref{sec:sepInhomP} (\emph{small} bandwidth), 
b) with a bit larger bandwidths  (\emph{moderate} bandwidth), and 
c) with much larger bandwidths (\emph{large} bandwidth).
\begin{table}[h!]
  \begin{center}
    \caption{
    Empirical significance level ($\gamma'=0$) based on $1000$ repetitions  of the permutation and $\chi^2$ tests for LGCP for different values of $\gamma''$, and at the significance level $\alpha = 0.05$. The global envelope test was performed with $999$ permutations of time using the $S(u,t)$-function evaluated at a $20\times20\times 20$ grid. The intensities were estimated by kernel smoothing using three different bandwidths, {\em small}, {\em moderate} and {\em large} presented in this order in the table for each case. 
The $\chi^2$-test was based on a $4\times 4\times 4$ grid.}
     \label{tab:3}
    \begin{tabular}{c|c|cc||c|c} 
    &&\multicolumn{2}{c||}{Bandwidths}&
    \multicolumn{1}{c|}{Global ERL envelope}
    &\\
         $\gamma''$ &$\beta_0$&$space$&$time$&
        $S(u,t)$ &$\chi^2$-test \\
        \hline\hline
        \multirow{3}{*}{0}&\multirow{3}{*}{5.05}
        &0.053&0.069&0.048&\multirow{3}{*}{0.042}\\
        &&0.1&0.12&0.047&\\
        &&0.15&0.17&0.049&\\
         \hline
        \multirow{3}{*}{0.5}&\multirow{3}{*}{5.00}
        &0.052&0.067
        &0.056&\multirow{3}{*}{0.059}\\
        &&0.1&0.12&0.062&\\
        &&0.15&0.17&0.057&\\
        \hline
        \multirow{3}{*}{1}&\multirow{3}{*}{4.9}
        &0.048&0.061
        &0.205&\multirow{3}{*}{0.126}\\
        &&0.1&0.11&0.15&\\
        &&0.15&0.16&0.12&
    \end{tabular}
  \end{center}
\end{table}
Table~\ref{tab:3} presents the empirical significance levels for the permutation test based on $S(u,t)$ with these bandwidth choices and for the $\chi^2$-test.
The following observations can be made from the empirical significance levels (case $\gamma'=0$):
\begin{itemize}
    \item For $\gamma''=0$ all empirical significance levels were fine.
        \item The empirical significance levels increased with increasing $\gamma''$. They were still close to the nominal level with $\gamma''=0.5$, but for $\gamma''=1$, all tests were liberal.
    \item For $\gamma''=1$, increasing the bandwidths of the permutation test based on $S(u,t)$ reduced the liberality of the test, but the nominal level was not reached for any bandwidth choice.
\end{itemize}
The first observation can be understood from the fact that $\gamma'=\gamma''=0$ implies the multiplicative form of the realisation of $\Lambda(u,t)$ as noted above. Thus, the additional clustering of the LGCP model caused by the $Z(u,t)$ with additive structure did not affect the empirical significance levels for $\gamma''=0$.
However, the non-additive form of $Z(u,t)$
for $\gamma''>0$ led to a non-multiplicative form of $\Lambda(u,t)$ and to liberal tests.
In fact, the bigger the $\gamma''$, the more severe the liberality.
The $\chi^2$-test was similarly liberal for $\gamma''>0$, due to unbalanced cell counts.

Thus, we conclude that the structure of the clusters has a major impact on the liberality of the tests for LGCPs and the considered first-order separability tests are truly valid only under the additive structure of $Z(u,t)$ (case $\gamma''=0$ of the considered model). Even though the nominal significance level was in the above experiment reached approximately also for $\gamma''=0.5$,
the liberality of the tests is further increased for any $\gamma''>0$ by increasing the values of the correlation parameters $\phi_1$ and $\phi_2$ or number of points controlled by the parameter $\beta_0$. 
\begin{table}[h!]
  \begin{center}
    \caption{The proportions of rejections among $1000$ repetitions of the permutation and $\chi^2$-tests for LGCP with $\gamma''= 0$ at the significance level $\alpha = 0.05$. The tests were performed as in Table~\ref{tab:3}.
    }
\label{tab:4}
   \begin{tabular}{c|c|cc||c|c}
    &&\multicolumn{2}{c||}{Bandwidths}&
    \multicolumn{1}{c|}{Global ERL envelope}
    &\\
         $\gamma'$ &$\beta_0$&$space$&$time$&
        $S(u,t)$ &$\chi^2$-test \\
        \hline\hline
        \multirow{3}{*}{0.5}&\multirow{3}{*}{5.05}
        &0.052&0.066&0.061&\multirow{3}{*}{0.044}\\
        &&0.1&0.12&0.058&\\
        &&0.15&0.17&0.062&\\
         \hline
        \multirow{3}{*}{1.5}&\multirow{3}{*}{4.7}
        &0.052&0.067
        &0.086&\multirow{3}{*}{0.078}\\
        &&0.1&0.11&0.15&\\
        &&0.15&0.17&0.172&\\
                \hline
        \multirow{3}{*}{3}&\multirow{3}{*}{4}
        &0.05 &0.06
        &0.266 & \multirow{3}{*}{ 0.163}\\
        & &0.1&0.11&0.343\\
        &  &0.15&0.16&0.45\\
                \hline
        \multirow{3}{*}{5}&\multirow{3}{*}{3.2 }
        &0.044&0.042 
        &0.49  &\multirow{3}{*}{ 0.242}\\
        &&0.094&0.092&0.595 &\\
        &&0.14&0.14&0.687&\\
     \end{tabular}
  \end{center}
\end{table}
Given these conclusions, we next explored the power of the tests for the LGCP model with $\gamma''=0$ for different values of $\gamma'$. The rejection rates among $1000$ simulations of the LGCP model are reported in Table \ref{tab:4}.
The results can be summarized as follows:
\begin{itemize}
\item As expected, the power of the tests increased with $\gamma'$, i.e.\ with increasing degree of non-separability.
    \item At a fixed level of $\gamma'$, the power of the permutation test based on $S(u,t)$ increased by increasing bandwidths, and this impression is more visible for large values of $\gamma'$.
\end{itemize}
The second observation is not surprising, because the non-separable structures of the model \eqref{eq:LGCP_m} were linear, leading to smoothly varying first-order intensities, which should be better captured by large bandwidths that involve larger amount of spatio-temporally correlated points in estimation of the intensity.

\section{Case study: UK 2001 epidemic foot and mouth disease data}\label{sec:fmd}
\begin{figure}[!htb]%
\begin{center}
\begin{tabular}{cc}
\includegraphics[width=0.31\textwidth, height=0.31\textheight]{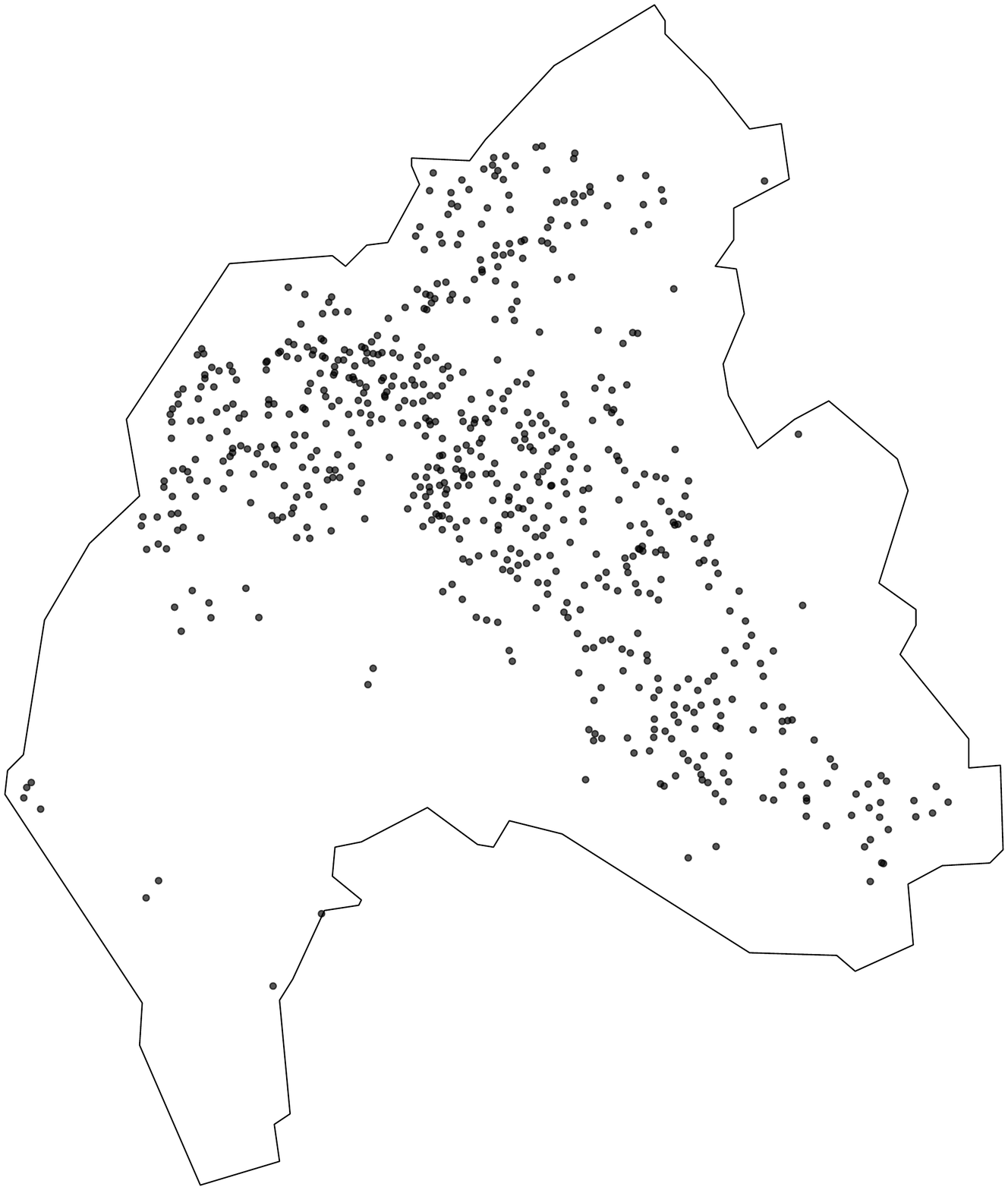}&
\includegraphics[width=0.35\textwidth, height=0.36\textheight]{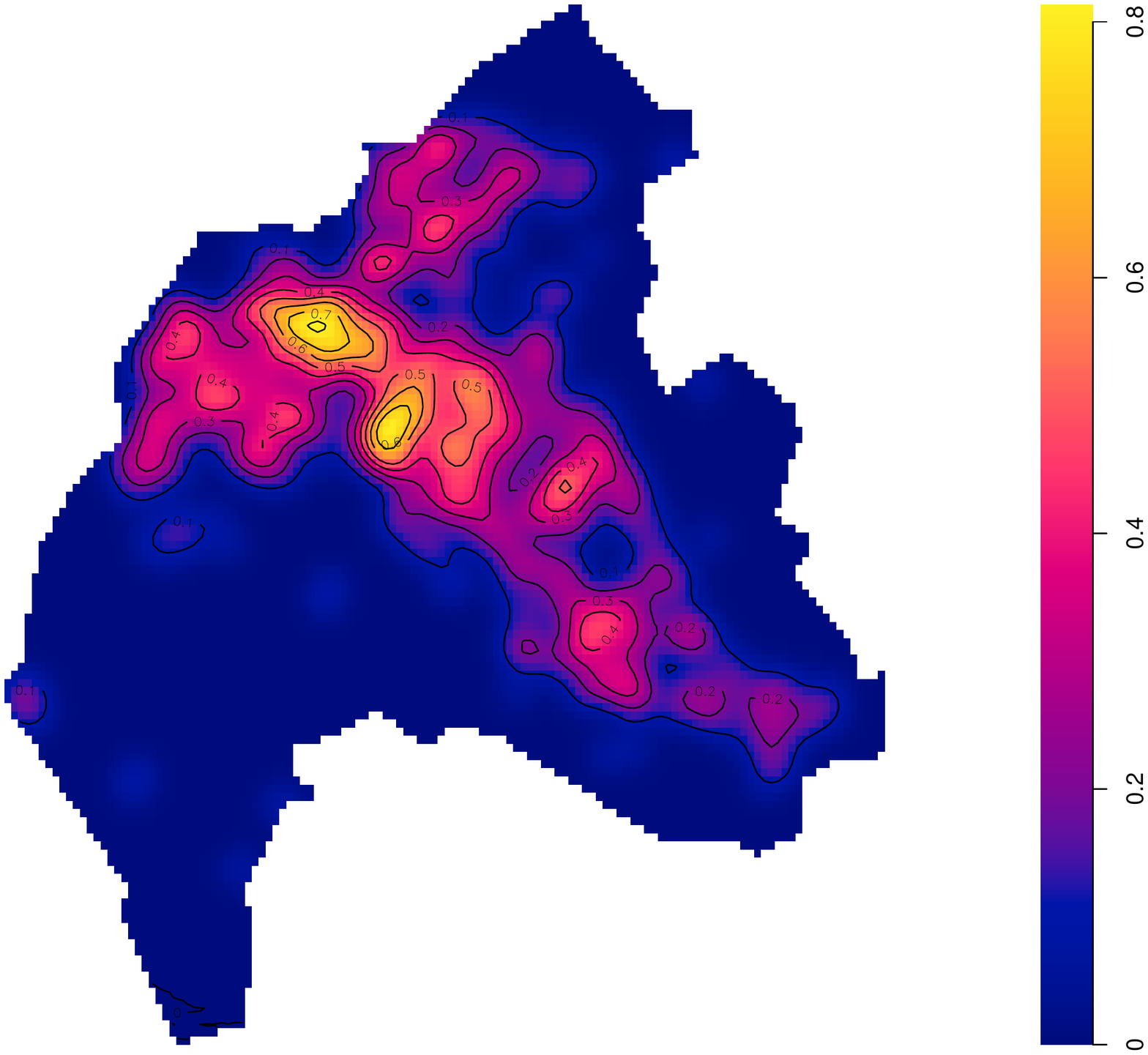}\\
 \multicolumn{2}{c}{\includegraphics[width=0.7\textwidth, height=0.3\textheight]{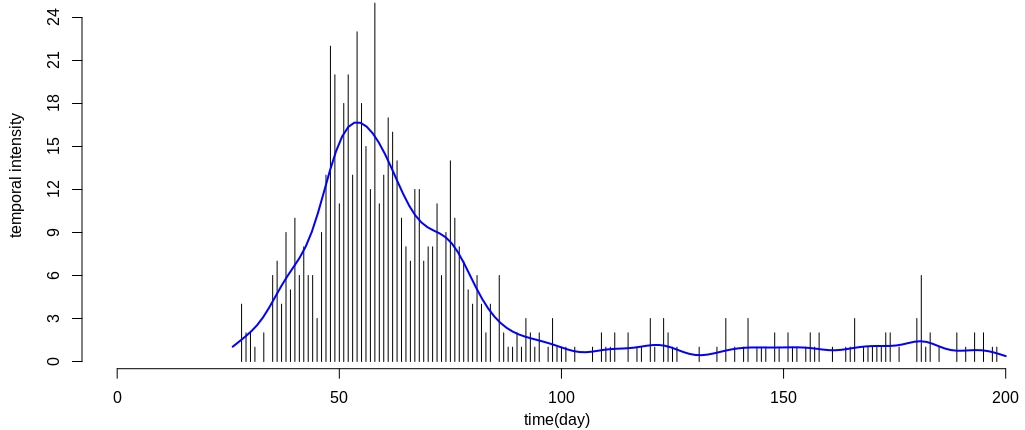}}
 \end{tabular}
\caption{Spatial point pattern of infected animals (top left panel),
$\hat\rho_{\mathrm{space}}(r)$ (top
right panel), and
$\hat\rho_{\mathrm{time}}(t)$ together with
the daily number of
infected animals (bottom panel) for the FDM data.}  \label{FMDPlot}%
\end{center}
\end{figure}

As a real data example, we investigated the first-order separability of the spatial and temporal components of the UK 2001 epidemic foot and mouth disease (FMD) dataset in Cumbria. This dataset was previously analyzed
in \citet{keelingetal:01}, \citet{Diggle:06,diggle:07}, \citet{gabriel:etal:18}, \cite{moeller:ghorbani:12} and \cite{ghorbani:13}.
For more information about the data see \cite{diggle:13}.
The data analyzed in this section
is taken from the R package {\em stpp} \citep{gabriel:etal:18}.
The area of Cumbria
is 5556.298 ${{\mathrm{km}}}^2$ and the data were collected for 200 days starting at February 1, 2001, so we let $T=[0,200]$.
Figure~\ref{FMDPlot} shows the spatial point pattern of 648 infected animals in the irregular region $W$ defined by Cumbria  (upper left panel),
and the daily number of infected animals (bars in the lower panel).
Further, the upper right and lower panels show  the estimated spatial intensity $\hat\rho_{{\mathrm{space}}}$ estimated by \eqref{eq:rhohat_space} with bandwidth
$b=1.83$ km and the estimated temporal intensity $\hat\rho_{{\mathrm{time}}}$ estimated by \eqref{eq:rhohat_time} with bandwidth of 3.86 days, respectively. The bandwidths were chosen by the same rules as in the simulation study (see Section \ref{sec:sepInhomP}).

Manifestly, the most incidents occur in the North-Western to South-Eastern (NW-SE) belt of Cumbria and within the first 100 days.
Closer inspection of the spatio-temporal data indicates that the epidemic has begun in the far north of the Cumbria and subsequently spread both south-west and south-east. Transmission of infection is thought to occur primarily between neighbouring farms but cases can also occur far from all pre-existing cases, possibly because of the unintended transport of infected material \citep{diggle:13}.
For the FMD dataset, what {\em is of interest} is how the farms’ locations collectively affect the progress of the epidemic in time. Is there any relationship between farms' location and  the times $t_i$ at which particular farms reported FMD cases? From statistical perspective, we want to see if the spatial and temporal components are independent.
In the following, we test the first-order separability of the fmd data, first assuming that the data follow an inhomogeneous Poisson process (Section \ref{sec:fmd_pois}) and second treating it as a clustered pattern (Section \ref{sec:SRMFMD}).

Both in the permutation and reconstruction method, 2499 simulations were generated under the null hypotheses and the global envelope test was performed for the empirical and simulated $S$-functions that were evaluated on a $50\times 50\times 10$ grid.
\subsection{Permutation and $\chi^2$ tests}\label{sec:fmd_pois}
We first tested the first-order separability of the FMD data by the permutation and $\chi^2$ tests, assuming that the data follows an inhomogeneous Poisson process.
The simple $\chi^2$-test was performed with the $3 \times 3 \times 3$ grid where the boundaries of the cells were determined using the 1/3- and 2/3- sample quantiles of the observed coordinates in each dimension, similarly to the experiments in Section~\ref{sec:performance}.
The grid had only 27 cells to ensure the observed counts in individual cells were high enough.
The resulting $p$-value was $6.6 \cdot 10^{-24}$, indicating very strong evidence against the null hypothesis of the first-order separability. 

\begin{figure}
\begin{center}
    \subfloat[Data]{\includegraphics[width=0.75\textwidth]{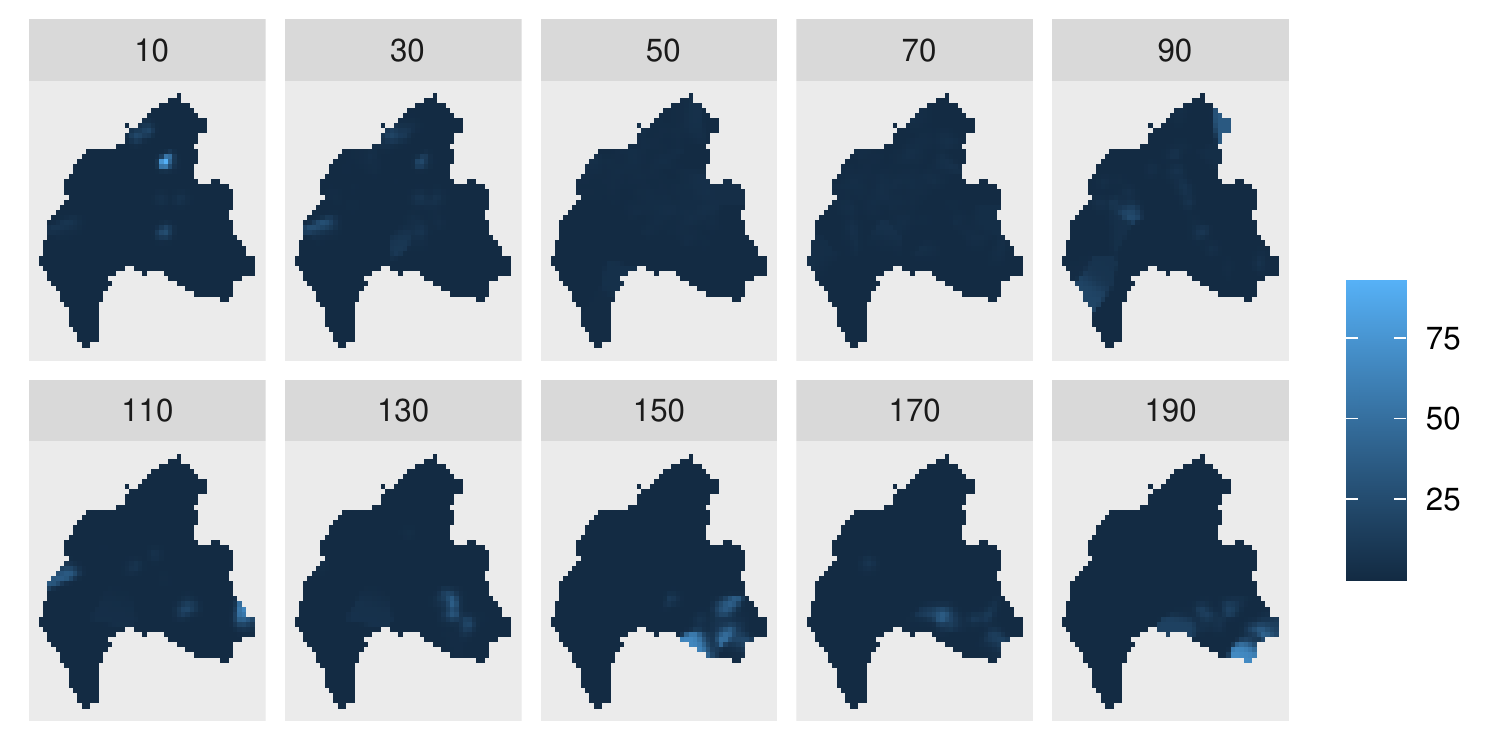}}\\
     \subfloat[Lower envelope]{\includegraphics[width=0.75\textwidth]{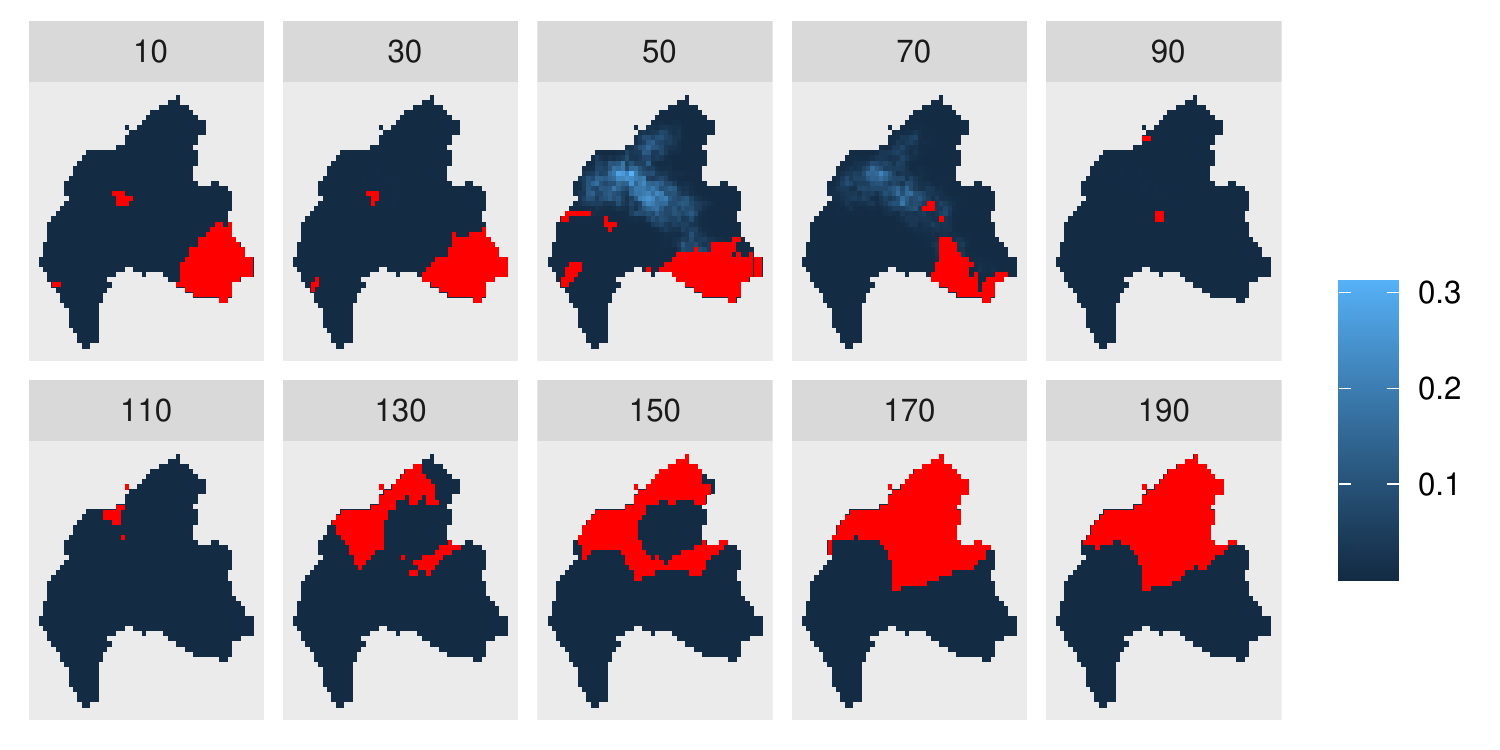}}\\
    \subfloat[Upper envelope]{\includegraphics[width=0.75\textwidth]{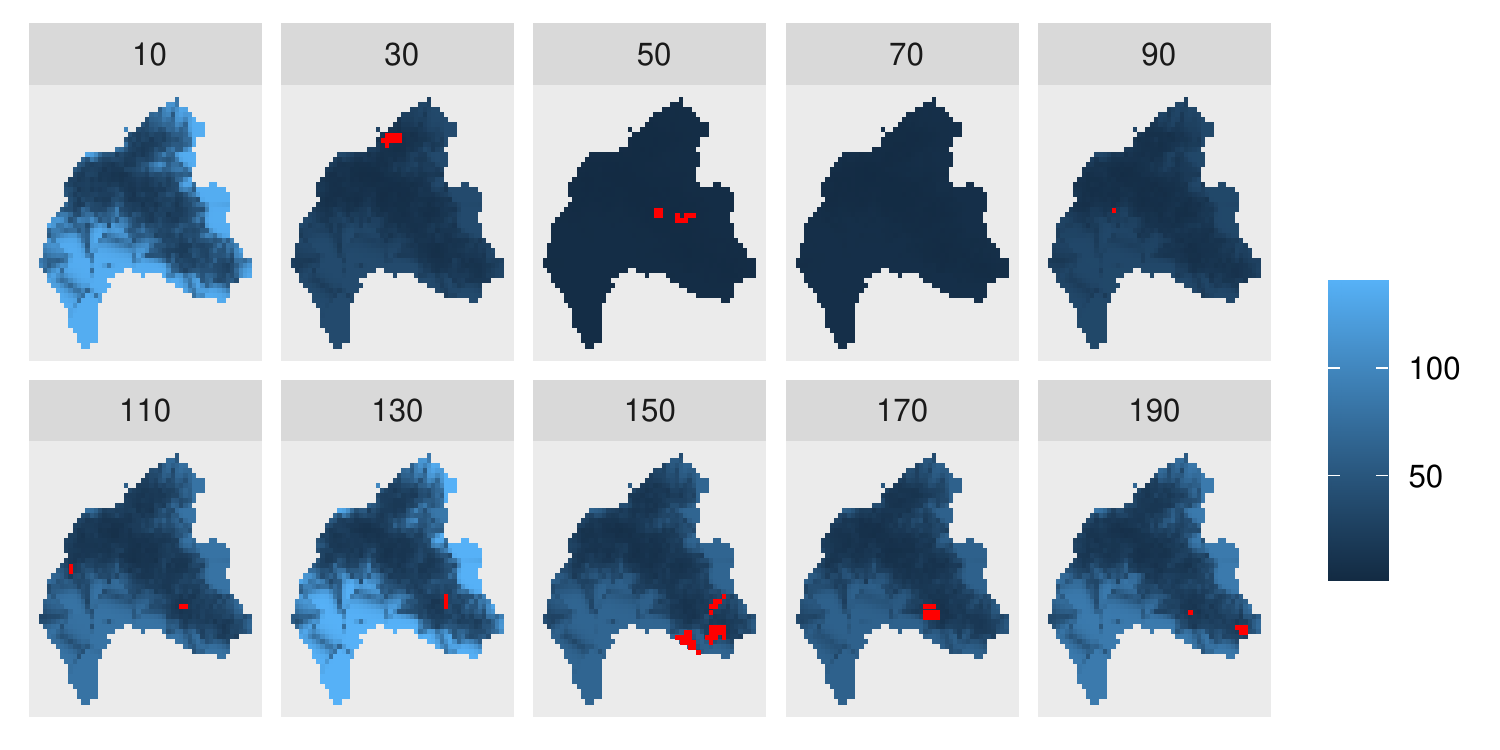}}
 \caption{Permutation based test: testing the first-order separability of the FMD data using the global ERL envelope test with the $S$-function ($p=4\cdot 10^{-4}$): (a) The empirical $S$-function  (blue color), (b) the lower envelope overlaid by the significant regions where the empirical function goes below the envelope, and (c) the upper envelope overlaid by the significant regions where the empirical functions goes above the envelope. The 95\% global envelope was constructed from 2499 simulations. The temporal coordinates of the grid are given above the corresponding plots.}\label{fig:S-FMD}
 \end{center}
\end{figure}

The output of the permutation based test is shown in Figure~\ref{fig:S-FMD}.
The test detects that the intensity of points was particularly high in the NW corner at time 30, in the middle of the region at time 50 and in the SE corner at later times 110-190. 
Simultaneously, the observed intensity goes below the lower envelope obtained from the permutations at quite large areas in the SE corner at early time and in the north and NE at later times (see description of the phenomena in Section \ref{sec:simexample}).

 
\subsection{Stochastic reconstruction method}\label{sec:SRMFMD}
For testing the first-order separability hypothesis in the FMD dataset, treating it as a clustered pattern, we applied the stochastic reconstruction approach described in Section~\ref{sec:SR}. 
To capture the interaction structure of the observed pattern $X$, 
we estimated the inhomogeneous space-time $K$-function where we used the non-separable kernel estimator of the intensity function $\hat{\rho}(u,t)$. We chose the non-separable estimator for this purpose since we are not sure whether the data were generated by a first-order separable process or not and we want to correctly estimate the interactions in both cases.
Furthermore we estimated the functions $\hat{D}_k$ to capture the arrangement of points in the clusters. The estimator of the intensity function to be used in the energy functional~\eqref{eq:energy} is the separable estimator $\hat{\rho}_{sep}(u,t)$ so that the separable first-order structure is enforced on the output patterns.

For tuning the algorithm, i.e.\ choosing the weights, upper bounds $T_K, R_K, T_D, R_D$ and the number $k_{max}$ in \eqref{eq:energy}, we used 
simulations from a model similar to the one fitted to the data in \citet{moeller:ghorbani:12}.
Namely, we considered a Poisson-Neyman-Scott type of process on the observation window $W \times T$ from the FMD dataset with inhomogeneous population of parents points following a Poisson process with the intensity function proportional to $\hat{\rho}_{sep}(X;u,t)$ and the mean number of parent points in $W \times T$ being 182. The mean number of offsprings per parent point was 3.5, their displacement around the parent point was governed by a trivariate Gaussian distribution with independent components and standard deviation 3.23 kilometers (in the spatial dimensions) and 7 days (in the temporal dimension), respectively.

As suggested in \citet{KonasovaDvorak}, from this model we generated 100 training patterns (to be used as inputs in the stochastic reconstruction procedure) and 999 testing patterns. For each training pattern we produced one output pattern using a given version of the stochastic reconstruction algorithm. To check the interaction structure of the output, we performed a global envelope test in which this output was treated as the observed data and the 999 testing patterns were treated as the Monte-Carlo replication. The test statistic was the inhomogeneous space-time $J$-function from \citet{CronieVanLieshout2015}. In this way we produced 100 $p$-values from the 100 training patterns. Uniformity of distribution of these $p$-values can be considered as an indication that the outputs of the stochastic reconstruction cannot be distinguished from the simulations from the true model, using this functional characteristic.

\begin{figure}[!htb]
\begin{center}
\begin{tabular}{c}
\includegraphics[width=0.5\textwidth]{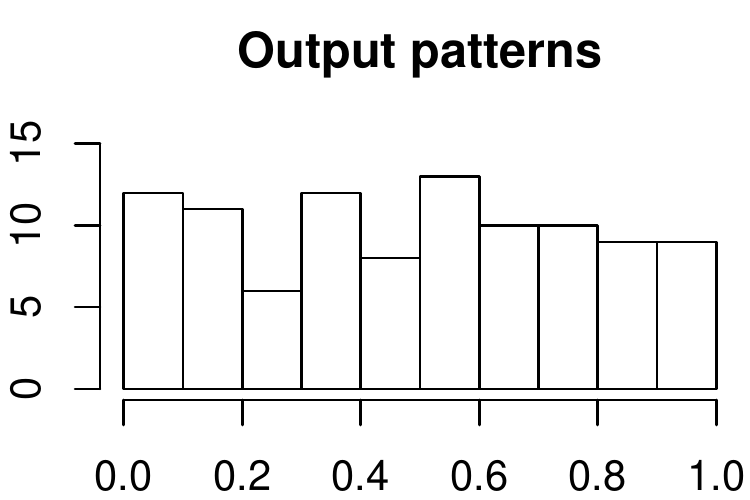}
\includegraphics[width=0.5\textwidth]{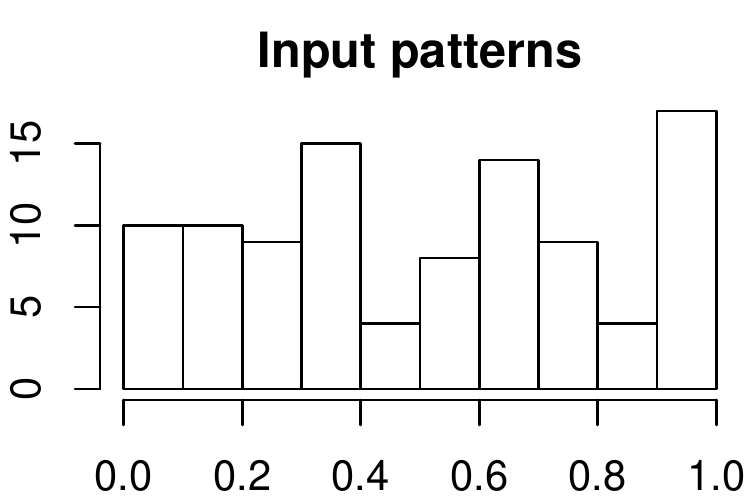}
 \end{tabular}
\caption{Histograms of the $p$-values obtained in the ERL global envelope test using the inhomogeneous $J$-function, for the outputs of the stochastic reconstruction procedure (left) and for comparison also for the respective input patterns (right). Uniformity of the distribution of $p$-values implies that the outputs cannot be distinguished from simulations from the true model using the inhomogeneous $J$-function.}\label{fig:SR}
\end{center}
\end{figure}

After some experimenting we chose the following values: $T_K=12$ days, $R_K=6$ km, $T_D=6$ days, $R_D=3$ km, $k_{max}=3$, $w_K=1$, $w_{D_k}=3\cdot10^3$, $w_\Delta=4\cdot10^5$.
The iterations were stopped either if 100 proposals were rejected in a row or the maximum number of $10^5$ iterations was reached. The weights were chosen so that, after the algorithm stops, the contribution of the individual terms in the energy functional is of the same order of magnitude. With these choices we performed the set of tests described in the previous paragraph, obtaining a very homogeneous population of $p$-values, see Figure~\ref{fig:SR} (left). For comparison we also performed the set of tests where the training patterns were treated as the observed data, obtaining again a rather uniform distribution of the 100 resulting $p$-values, see Figure~\ref{fig:SR} (right). We conclude that with these choices the stochastic reconstruction algorithm produces outputs which correctly reproduce the interaction structure of the given model, and the separable form of the first-order structure is enforced by the choice of the energy functional. Hence the outputs can be used in a Monte-Carlo test of the first-order separability hypothesis in place of independent simulations from the correct model.

\begin{figure}
\centering
    \subfloat[Data]{\includegraphics[width=0.75\textwidth]{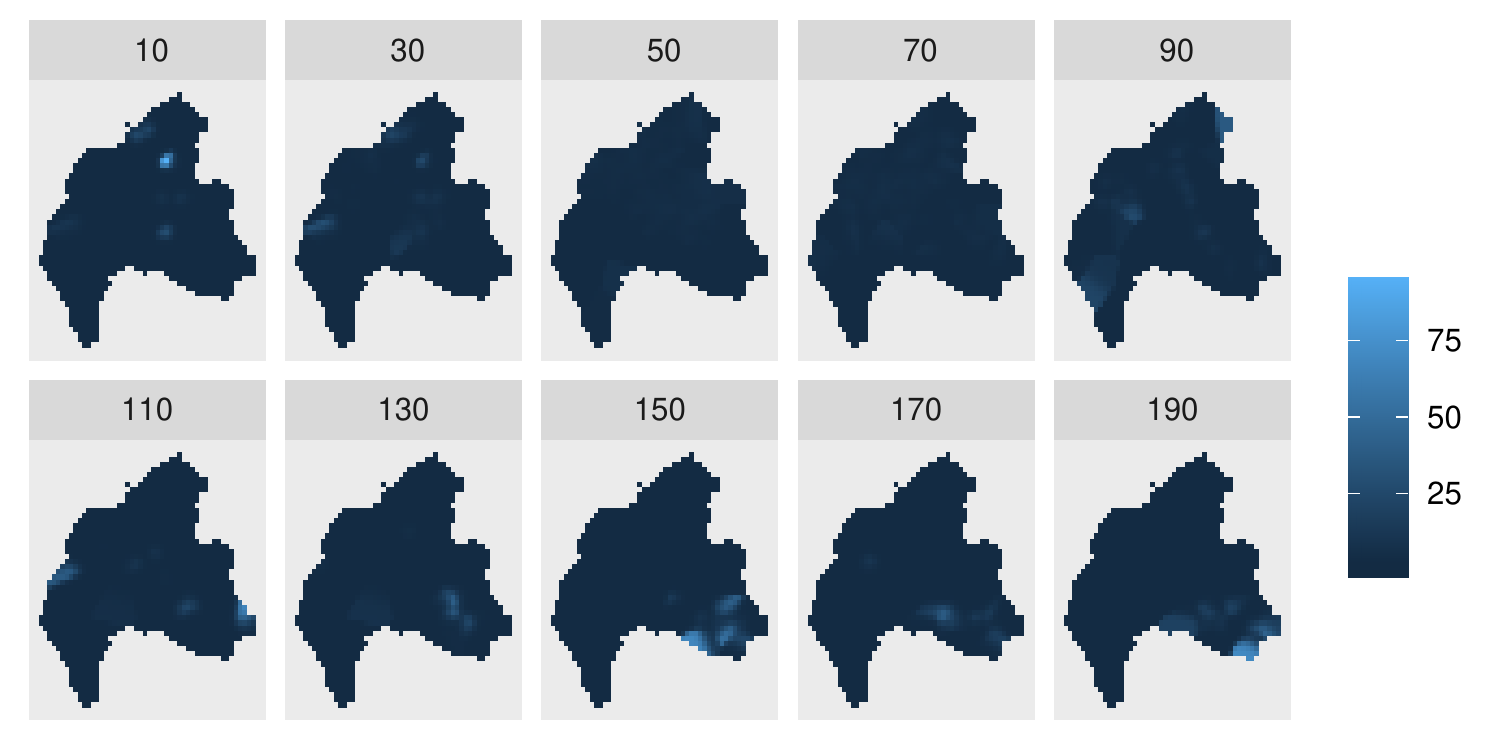}}\\
     \subfloat[Lower envelope]{\includegraphics[width=0.75\textwidth]{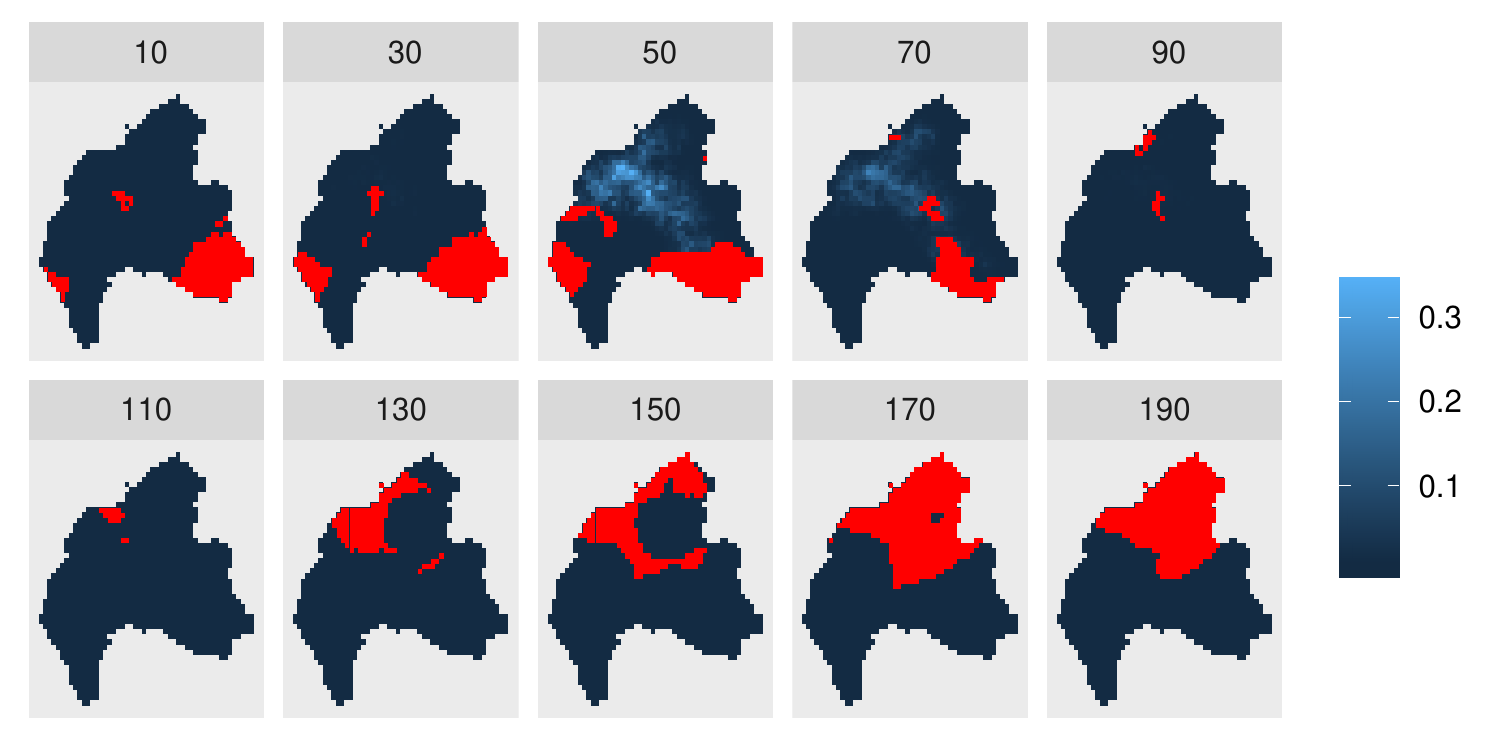}}\\
    \subfloat[Upper envelope]{\includegraphics[width=0.75\textwidth]{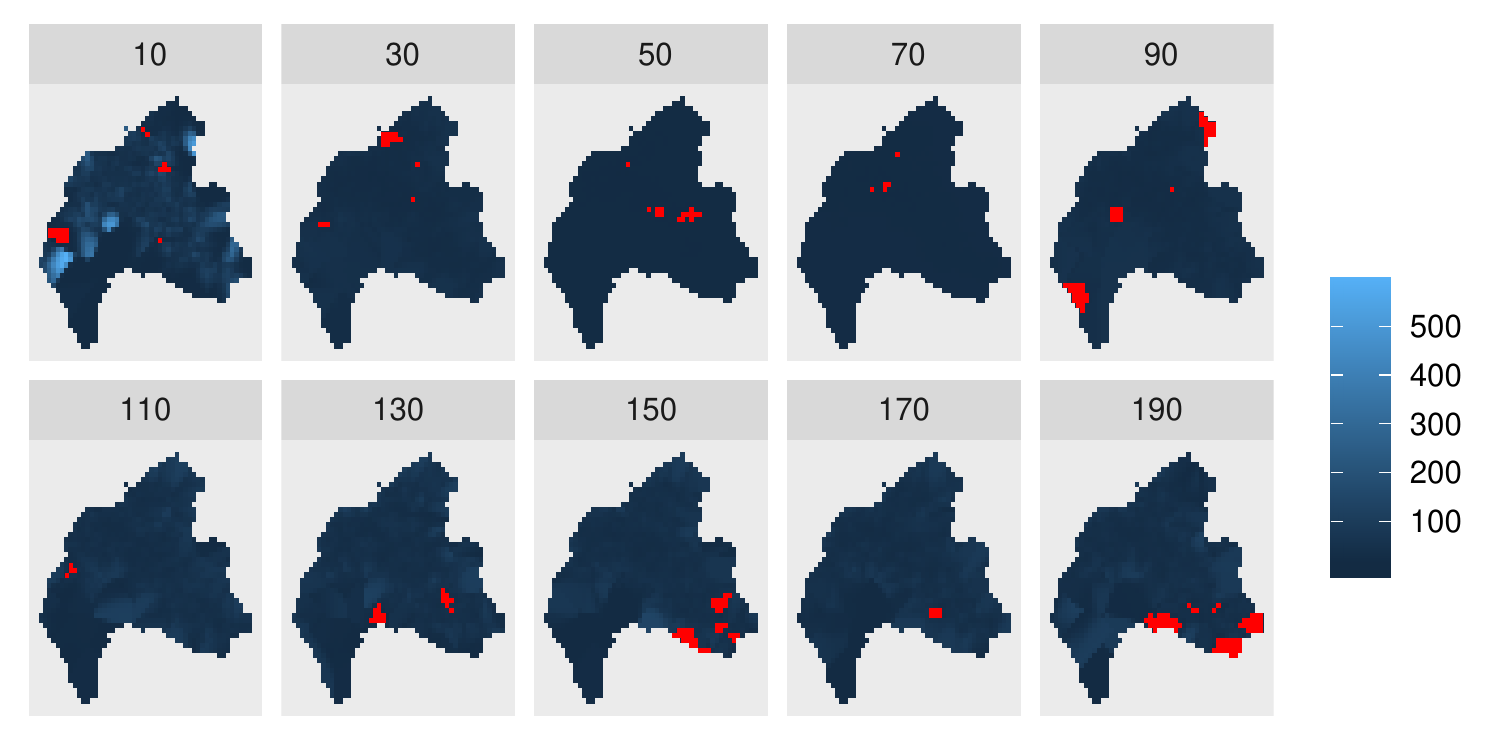}}\\
\caption{Stochastic reconstruction method: testing the first-order separability of the FMD data using the global ERL envelope test with the $S$-function ($p=4\cdot 10^{-4}$): (a) The empirical $S$-function (blue color),
(b) the lower envelope overlaid by the significant regions where the empirical function goes below the envelope, and (c) the upper envelope overlaid by the significant regions where the empirical functions goes above the envelope. The 95\% global envelope was constructed from 2499 simulations. The temporal coordinates of the grid are given above the corresponding plots.}\label{fig:SR_FMD}
\end{figure}

With the choices above we produced 2499 reconstructions of the FMD dataset and used these to test the first-order separability hypothesis as described in Section~\ref{sec:permutation-test}, using the function \eqref{eq:Sfun}. We employed the ERL version of the global envelope test at the significance level of 5~\%. The test rejected the null hypothesis with the smallest possible p-value $1/2500 = 4\cdot10^{-4}$. The outcome of the test, together with the significant regions where the data function lies above/below the envelope, is given in Figure~\ref{fig:SR_FMD}.

The significant regions clearly indicate that the epidemic has shifted over time, starting in the north-western part of the region and gradually moving to the south-eastern part (this can be also seen from the interactive plots in the accompanying website \url{http://msekce.karlin.mff.cuni.cz/~dvorak/software/STseparability.html}). This is not consistent with the null hypothesis of the first-order separability and the test acknowledges this issue by indicating that, compared to the reconstructed patterns with the separable first-order structure, the observed pattern has some missing points in the early times in the south-eastern part and in the later times in the north-western part (see the lower envelope in Figure~\ref{fig:SR_FMD}). Similarly, some excess points are present in the observed pattern at various locations and times (see the upper envelope in Figure~\ref{fig:SR_FMD}). It is interesting to note that this effect is strong enough not to be confused with clustering and that the significant regions are very similar to those reported by the permutation-based tests in Figure~\ref{fig:S-FMD}.

\section{Discussion and conclusion}
In the analysis of  spatio-temporal point patterns, modelling of the intensity function which characterizes the first-order structure is one of the first steps.
However, modelling the joint distribution of the spatial locations and time of occurrences of a STPP can be challenging due to the curse of dimensionality.
Hence, in the literature on spatio-temporal point processes, either for modelling purpose or for checking second-order separability hypothesis, first-order separability  is usually assumed \citep{Gabriel:Diggle:09,moeller:ghorbani:12}. This assumption allows us to express  the spatio-temporal intensity in a multiplicative form which nicely ease both modelling and estimation. However, in practical applications, the separability assumption can be quite restrictive and unrealistic and should be tested in the early stage of the analysis.

Heuristically, for any infinitesimal region $(du\times dt)\subseteq\R^2\times \R$, $(u,t)\in (du\times dt)$, with (Lebesgue) volume $|du\times dt|=\de u\times \de t$, the spatio-temporal intensity function of a STPP can be described by
\begin{align*}
\rho(u,t)\, \de u \, \de t
=\mu(du \times dt)
=\mathbb{P}(|X\cap (du\times dt)|=1).
\end{align*}
From this perspective, the first-order separability can be considered as the first-order spatio-temporal independence and therefore, in principle, one may think about a connection between the separability of the intensity function and the independence of two random variables. 
The independence of two random variables can be tested based on the distance between empirical cumulative distribution function \citep{blum:etal:61}
\begin{align*}
\int\int\left[F_n(u,v)-F_n(u)F_n(v)\right]^2 \, \de F(u, v).
\end{align*}
In fact, a similar test statistic could be constructed for the first-order separability test of a STPP as well. Namely,
considering the separable intensity \eqref{eq:rhoSep}, for Borel sets $A \subseteq \mathbb{R}^2$ and $B \subseteq \mathbb{R}$,  it is natural to expect that the quantity
\begin{align*}
\left[\hat \mu(A\times B)-\hat\mu_{sep}(A\times B)\right]^2=
\left[\int_{A\times B}\hat\rho(u,t)\, \de (u,t)-\frac1n \int_A\hat\rho_{{\mathrm{space}}}(u)\, \de u\int_B\hat\rho_{{\mathrm{time}}}(t)\, \de t\right]^2
\end{align*}
is close to zero under the first-order separability hypothesis. 
This quantity can be used to assess the separability of the intensity function locally in a given sub-region $A\times B$. 
To obtain a  test statistic for global assessment, one should sum over disjoint sub-regions $A_i\times B_j$, $i=1,\ldots, k$, $j=1,\ldots, l$, where $k$ and $l$ denote numbers of quadrats in space and time, respectively. In particular, if $A=(0, u]$ and $B=(0,t]$, then the test of first-order separability can be based on the distance between kernel estimates of intensity measures,
\begin{align}\label{eq:testSepGA}
\int\int\left[\hat \mu\big((0,u]\times(0,t]\big)-\hat\mu_{sep}\big((0,u]\times(0,t]\big)\right]^2 \, \de u \, \de t.
\end{align}
This idea has  been roughly stated in \citet[page 451]{Diggle:Gabriel:10}.
In fact, in the results not shown, motivated by this characterization, we explored a test function
\begin{align}\label{eq:Sfun2}
S^*(u,t)= \left( \hat\rho(u,t) - \hat\rho_1(u)\hat\rho_2(t) \right)^2
=\left(\hat\rho(u,t) - \hat{\rho}_{{\mathrm{space}}}(u)\hat{\rho}_{{\mathrm{time}}}(t)/n \right)^2,\quad  (u,t)\in W\times T,
\end{align}
which however did not perform as well as the chosen statistic \eqref{eq:Sfun}, leading to lower power in our examples and loosing the comparison of the order of the separable and non-separable intensities.
Our test statistic \eqref{eq:Sfun} is instead based on the quotient of non-parametric estimates of the non-separable and separable intensity functions.
In fact, the non-separable intensity plays the main role in our test statistic; the separable intensity is constant under permutations, and thus serves only as a scaling factor.
We note that the $\chi^2$-test is based on the counts of points in joint sub-regions, but also based on dividing the separable and non-separable quantities. It requires that the observed counts in individual cells be at least five to make the $\chi^2$ approximation applicable, otherwise one could use the Fisher's exact test that we have not considered in this paper.

We proposed permutation and $\chi^2$ tests which are appropriate to work under the Poisson assumption.  
The permutation test requires the permutations and calculations of the test statistic for each permutation. While permutations are computationally cheap, estimation of the spatio-temporal intensities can take some time. As a reward, one obtains however a graphical test that shows the spatial areas and times where the data contradict the null hypothesis.
On the other hand, the $\chi^2$-test is a simple and computationally cheap alternative.
Both tests turned out to have a good performance also for processes with weak clustering. 
Hence, for Poisson and weakly clustered processes, we recommend the $\chi^2$-test  as a fast preliminary test, and the permutation test based on the test function \eqref{eq:Sfun} in order to learn where and when the potential non-separable structures occur.

For other processes, stochastic reconstruction method was proposed. It is a computationally expensive method, whose implementation for a specific data requires some experimenting. However, it can also provide detailed information about the possible non-separabilities in the intensity of the process.

One challenging point in the use of our test statistics can be their sensitivity to the choice of bandwidth. We believe that the choice of the bandwidths in intensity estimation can affect the performance of the tests and thus their proper choice can definitely increase the power of the tests. For Poisson processes the procedures for bandwidth selection as explained in Subsection~\ref{sec:sepInhomP} can be used. For non-Poisson processes  there is unfortunately no automatic way for bandwidth selection.
In general, the user should have some prior information on the smoothness of the first-order properties in the data and choose the bandwidth with respect to this understanding. This is  important for the stochastic reconstruction approach, and also for the permutation based test if applied to data with small scale clustering.

\section*{Acknowledgements}
The authors are grateful to Ottmar Cronie for sharing code of the $J$-function, and thank Mikko Kuronen for discussions on the topic and implementations. 
MG was financially supported by the Kempe foundations (SMK-1750),
MM and NV by the Academy of Finland (Project numbers 295100, 306875 and 327211), and JD
by the Grant Agency of the Czech Republic (Project No.\ 19-04412S). 
The work was mainly done when NV was visiting Luke.
The authors wish to acknowledge CSC - IT Center for Science, Finland, for computational resources.
\bibliographystyle{dcu}
\bibliography{papertsh}
\end{document}